\documentclass[11pt,a4paper]{article}
\usepackage{jheppub}

\usepackage{slashed}
\usepackage{hyperref}
\usepackage{epsfig,amsmath,amsfonts,amssymb,amsthm}
\usepackage[normalem]{ulem}
\usepackage{color}
\usepackage{array}
\usepackage{multirow}
\usepackage{subfig}
\usepackage{wrapfig}
\usepackage{graphicx}
\usepackage{tabu}

\usepackage{slashed}
\usepackage{float}

\usepackage[justification=centering]{caption}

\usepackage{booktabs}
\usepackage{comment}
\usepackage{enumitem}
\usepackage{graphicx}
\usepackage{xcolor}
\usepackage{verbatim}
\usepackage{dsfont}
\setlist[itemize]{leftmargin=*}

\usepackage{comment}
\usepackage{tikz}
\usetikzlibrary{decorations.pathmorphing,decorations.markings}

\tikzset{
  psi/.style={
    decoration={
      markings,
      mark=at position 0.6 with {\arrow{>}}
    },
    postaction={decorate},
    double,
    double distance=1pt
  },
  psiNoArrow/.style={
    decoration={
      markings,
      mark=at position 0.6 with 
    },
    postaction={decorate},
    double,
    double distance=1pt
  },
  nucleon/.style={
    decoration={
      markings,
      mark=at position 0.6 with {\arrow{>}}
    },
    postaction={decorate}
  },
  external/.style={},  %{circle,fill,inner sep=1.5pt}
  gluon/.style={
  decorate, draw=black, 
  decoration={coil,amplitude=4pt, segment length=5pt}
  },
  particle/.style={draw=black, postaction={decorate}, decoration={markings,mark=at position .5 with {\arrow[draw=black]{>}}}},
 photon/.style={decorate, decoration={snake,amplitude=2pt, segment length=5pt}, draw=black}
}

\newcommand{\be}{\begin{equation}}
\newcommand{\ee}{\end{equation}}
\newcommand{\bea}{\begin{equation}\begin{aligned}}
\newcommand{\eea}{\end{aligned}\end{equation}}

\newcommand{\GeV}{\text{GeV}}

\newcommand{\hc}{\text{h.c.}}

\newcommand{\gsim}{\lower.7ex\hbox{$\;\stackrel{\textstyle>}{\sim}\;$}}
\newcommand{\lsim}{\lower.7ex\hbox{$\;\stackrel{\textstyle<}{\sim}\;$}}

\definecolor{grey}{cmyk}{0,0,0,0.75}
\definecolor{tangerine}{cmyk}{0,0.5,1,0}
\definecolor{darkgreen}{cmyk}{1,0,1,0.23}
\definecolor{Red}{rgb}{1,0,0}
\definecolor{Blue}{rgb}{0,0,1}
\definecolor{Green}{rgb}{0,1,0}
\definecolor{Grey}{cmyk}{0,0,0,0.75}
\definecolor{Tangerine}{cmyk}{0,0.5,1,0}
\definecolor{Darkgreen}{cmyk}{1,0,1,0.23}
\definecolor{Cyan}{cmyk}{1,0,0,0}
\definecolor{Yellow}{cmyk}{0,0,1,0}
\definecolor{darkblue}{cmyk}{1,0.69,0,0.11}

\newcommand{\nicpb}{Laboratory of High Energy and Computational Physics, NICPB, R\"avala pst. 10, 10143 Tallinn, Estonia}
\newcommand{\hki}{Department of Physics and Helsinki Institute of Physics, University of Helsinki, Gustaf H{\"a}llstr{\"o}min katu 2, {FI-00014} University of Helsinki, Finland}
\newcommand{\sumb}{Department of Physics, SEAS, Bennett University, Greater Noida, Uttar Pradesh -201310, India}

\begin{document}

\title{On the family discrimination in 331-model}

\author[a]{Katri Huitu,}
\author[b]{Niko Koivunen,}
\author[b]{Timo  K\"arkk\"ainen,}
\author[c]{Subhadeep Mondal}
\affiliation[a]{\hki}  
\affiliation[b]{\nicpb} 
\affiliation[c]{\sumb}

\emailAdd{niko.koivunen@kbfi.ee, subhadeep.mondal@bennett.edu.in}

\abstract{
In the so-called 331-models the gauge anomalies cancel only if there are three generations of fermions. This requires one of the quark generations to be in a different representation than the other two. But which generation is treated differently?
In this work we study how the choice of differently treated generation effects the quark flavour structure and how
%if
the discriminated generation can be deduced from experiments. We study a general model based on $\beta=-1/\sqrt{3}$, which contains exotic quarks with same electric charges as SM quarks. We take fully into account the effects from exotic quark mixing with the SM quarks, which is often omitted in literature.
We will also pay particular attention to $125$ GeV Higgs, and show analytically why its flavour violating couplings between SM quarks are suppressed.

}

\maketitle

%%%%%%%%%%%%%%%%%%%%%%%%%%%%%%%%%%%%%%%%%%%%%%%%%%%%%%%%%%%%%%%%%%%%%%%%%%%%%%%%%%%%
%%%%%%%%%%%%%%%%%%%%%%%%%%%%%%%%%%%%%%%%%%%%%%%%%%%%%%%%%%%%%%%%%%%%%%%%%%%%%%%%%%%%%%%%
\section{Introduction}
The Standard model (SM) extensions based on $SU(3)_c\times SU(3)_L\times U(1)_X$ gauge symmetry are called 331-models. The appealing feature of 331-models is their ability to explain the number of fermion families in nature. 
This follows from cancellation of gauge anomalies, which differs crucially from the SM. In the SM the pure $SU(2)_L$-anomaly cancels automatically due to properties of $SU(2)_L$-generators, $\{\sigma^a/2,\sigma^b/2\}=\delta^{ab}/2$, regardless of the particle content. In the 331-model the pure $SU(2)_L$-anomaly is replaced with $SU(3)_L$-anomaly. The $SU(3)_L$-generators do not share the special property of $SU(2)_L$-generators, and the pure $SU(3)_L$-anomaly does not cancel automatically. Unlike in the SM, the particle content has to be arranged in specific way for $SU(3)_L$-anomaly to cancel. 

In 331-models the left-handed fermions are traditionally assigned into triplets and antitriplets. The minimal particle content is achieved by demanding only one triplet/antitriplet per SM fermion doublet. In this way there is one new fermion per triplet, compared to SM. With this assumption the $SU(3)_L$-anomaly only cancels if the number of fermion triplets is same as the number of antitriplets. 
This fixes the number of generations to be integer multiple of three. However, the QCD looses asymptotic freedom if the number of generations is larger than four. By demanding asymptotic freedom, therefore, the only remaining possibility is to have three generations.
In 331-model the gauge anomalies cancel between generations, instead of within generation, like in SM\footnote{It is possible to construct 331-models where the gauge anomalies cancel within a generation, by allowing for components of SM fermion douplets to be distributed into multiple triplets/antitriplets \cite{Deppisch:2016jzl, Sanchez:2001ua} . In these sequential  models the number of generations is not predicted.}. 

Taking into account the colour charge, there are nine quark triplets/antitriplets and three lepton triplets/antitriplets. One therefore has to have six triplets and six antitriplets for the anomalies to cancel if all the lepton generations are assigned into triplets. Therefore one of the quark generations has to be placed in a triplet and two in antitriplets. One of the quark generations is then treated differently in 331-models. 
Question is, which generation?

The unavoidable consequence of one of the quark generations being in different representation is appearance of quark flavour changing neutral currents (FCNC) at tree-level \cite{Glashow:1976nt, Paschos:1976ay}. 
All the neutral bosons, except photon and gluon, mediate quark flavour violating processes at tree-level. The lepton generations are in the same representation and there are no flavour violation in the lepton sector. 
In the SM the neutral current flavour changing processes are forbidden at tree-level due to Glashow-Iliopoulos-Maiani (GIM) mechanism. Therefore these processes are very sensitive to new physics.
The choice of the discriminated quark generation leads to different flavour violating structures for quarks. In this work we study the effect of the choice of the discriminated generation in neutral meson mixing, which provides the most stringent constraints on the flavour violation structure of quarks.

The third generation is sometimes advocated to be the discriminated generation on the grounds of explaining the relative heaviness of the top quark compared to the other SM quarks \cite{Frampton:1994rt}. This is because the third generation will get its mass from different electroweak scale vacuum expectation value (VEV) than the 1st and 2nd, allowing VEV to be tuned to make top quark heavy. However, this does not explain the hierarchy between other SM quarks, and the bottom quark would also become heavy with this reasoning. The VEVs are not enough to explain the whole mass hierarchy and therefore hierarchy must exist in the Yukawa couplings themselves. The quark family discrimination has been studied in the past in 
\cite{GomezDumm:1993oxo,Liu:1994rx,Long:1999ij,Rodriguez:2004mw} and more recently in \cite{Oliveira:2022vjo}.
The consensus in the literature is that the discriminated third generation yields the smallest contribution to the flavour changing quark processes.

In the context of 331-model, the FCNCs are studied in two different ways in the literature: either assuming that the quark rotation matrices have some hierarchy originating from some preset structure in the quark mass matrices and deriving bound on the $SU(3)_L$-breaking scale \cite{Long:1999ij,Rodriguez:2004mw}, or by assuming nothing about the quark rotation matrices and deriving the constraints on the quark rotation matrix elements \cite{GomezDumm:1993oxo,Liu:1994rx,Promberger:2007py}. We opt to follow the former route in this article.    
Most common ansatz for quark mass matrix structure is the Fritzsch ansatz \cite{Fritzsch:1977vd, Fritzsch:1979zq}, used in the analysis by \cite{GomezDumm:1993oxo,Long:1999ij, Rodriguez:2004mw,Cabarcas:2007my,Martinez:2008jj}. 
In Fritzsch ansatz the off-diagonal elements of quark rotation matrices $U$, follow $U_{ij}\sim(m_i/m_j)^{1/2}$, which does not fit to measurements of Cabibbo-Kobayashi-Maskawa (CKM) matrix elements any more.  Modifications to Fritzsch ansatz in non-331 context are studied in \cite{Belfatto:2023qca}.
Other textures are used in 331 context in  \cite{Martinez:2008jj, CarcamoHernandez:2005ka}. 
These textures take into account only the mixing among SM quarks and are not applicable in the full mass matrix if exotic quarks are mixing with SM ones.

The 331-models have freedom in the definition of the electric charge, due to $SU(3)_c\times SU(3)_L\times U(1)_X$ gauge group having one additional diagonal generator compared to the SM. The electric charge corresponds to a linear combination of diagonal generators and in a generic 331-model it is
\begin{equation}\label{electric charge}
Q=T_3+\beta T_8+X,
\end{equation}
where the $T_3$ and $T_8$ are the diagonal $SU(3)_L$ generators, $X$ the $U_X$-charge and the parameter $\beta$ is a free parameter. Most studied models are based on $\beta=\pm 1/\sqrt{3}$ \cite{Georgi:1978bv,Singer:1980sw, Valle:1983dk, Montero:1992jk, Foot:1994ym, Long:1995ctv, Long:1996rfd, Pleitez:1994pu, Long:1997vbr,Dong:2013ioa} and $\beta=\pm \sqrt{3}$ \cite{Pisano:1992bxx, Frampton:1992wt, Foot:1992rh, Tonasse:1996cx, Nguyen:1998ui}.  
The models based on  $\beta=\pm\sqrt{3}$ have a complicated scalar sector as they traditionally include three scalar triplets and a scalar sextet in order to give mass to all of the charged leptons. The scalar sector of the traditional  $\beta=\pm1/\sqrt{3}$ models consist of only three scalar triplets and is simpler. We will be working with this simpler option. The models based on $\beta=\pm 1/\sqrt{3}$ contain new quarks (which we dub exotic quarks) which have same electric charges as SM quarks, which in general mix with the SM quarks. The models with $\beta= -1/\sqrt{3}$, the value we use in this work, contain one exotic up-type quark and two exotic down-type quarks\footnote{The models based on $\beta= 1/\sqrt{3}$ contain two exotic up-type quarks and one exotic down-type quark.}. As a result of this the most general up-type and down-type quark mass matrices are $4\times 4$ and $5\times 5$, respectively, with all elements non-zero. The full mixing between SM quarks and exotic quarks is rarely taken into account in the literature, but considered for example in \cite{Benavides:2009cn}. We will fully take into account the mixing between SM quarks and exotic quarks by using quark mass matrix textures inspired by the structure that appears in the Froggatt-Nielsen (FN) mechanism \cite{Froggatt:1978nt}. We will, however, remain agnostic about the origin of the mass matrix, and do not assume any underlying flavour symmetries, like in FN mechanism.

By using these Froggatt-Nielsen inspired quark mass matrix textures, we study the affect of the choice of the discriminated quark generation on the flavour violating couplings, and use the neutral meson mixing to place bounds on the $SU(3)_L$-breaking scale.  
The use of FN-like textures allows us to obtain analytical understanding on the magnitude of the flavour violating couplings, which makes the interpretation transparent.   
We will also pay particular attention on the flavour changing couplings of $125$ GeV Higgs and provide analytical understanding on their suppression, which has been largely omitted in the literature. 

The paper is structured as follows. In Section \ref{particle content} we present the particle content of the model. The scalar masses and eigenstates are presented in \ref{scalar masses}. In Section \ref{quark yukawa couplings and masses} we present the Yukawa sector of quarks and the assumptions we make about the structure of their mass matrices. In the section \ref{neutral meson mixing} we compute the contribution to the neutral meson mass difference and present the order of magnitude estimates of flavour violating couplings for all FCNC messengers, based on the assumptions made in Section \ref{quark yukawa couplings and masses}. We compare the results from different quark generation assignments in Section \ref{numerics}. Finally we conclude in Section \ref{conclusions}.

\section{Particle content}\label{particle content}
We study 331-model with $\beta=-1/\sqrt{3}$, where the beyond the Standard model (BSM) particles don't have electric charges not already present in the SM. The electric charge operator \eqref{electric charge} then is 
\begin{equation}
Q=T_3+\beta T_8+X=T_3-\frac{1}{\sqrt{3}}T_8+X.
\end{equation}
We assume minimal particle content that provides masses to electrically charged fermions at tree-level, that is three scalar triplets. 

\subsection{Fermion representations}\label{fermion representations}
The left-handed leptons are assigned to $SU(3)_{L}$ -triplets and the right-handed charged leptons are assigned to $SU(3)_{L}$-singlets:
\begin{eqnarray}
&&L_{L,i}=\left(\begin{array}{c}
\nu_{i}\\
e_{i}\\
\nu'_{i}
\end{array}
\right)_{L}\sim (1,3,-\frac{1}{3}),\quad 
e_{R,i}\sim (1,1,-1),\quad i=1,2,3.
\end{eqnarray}
%The numbers in the parantheses label the transformation properties under the gauge group  $SU(3)_{c}\times SU(3)_{L}\times U(1)_{X}$. 
The $\nu'_{L,i}$ are new leptons with zero electric charge. This neutrino sector is not realistic, as one of the neutrinos is left massless and the other two mass degenerate at tree-level. Loop corrections are required to lift the degeneracy of the neutrinos \cite{Valle:1983dk}. The neutrino sector in this type of model requires to be extended to make it compatible with the neutrino data.  For example, additional right-handed neutrino singlets have been studied in a similar model \cite{Huitu:2019mdr}. Regardless, the neutrino sector is not related to the problem we are studying in this work and we leave it as it is.  
Since the charged leptons are all placed in the same representation, the charged lepton mass matrix is diagonalized simultaneously with the Yukawa matrix. Therefore, there is no charged lepton flavour violation.  

Equal number of triplets and antitriplets are required for gauge anomalies to be cancelled.
Since all the left-handed leptons are in triplets, one of the quark families has to be placed in a triplet (due to three colours, it corresponds to three triplets), and the other two in antitriplets. We will consider all the three different choices for the discriminated quark generation. Below we present the choice of placing the first generation in the triplet,
\begin{equation}
Q_{L,1}=\left(\begin{array}{c}
u_{1}\\
d_{1}\\
U
\end{array}
\right)_{L}\sim (3,3,\frac{1}{3}),\quad 
Q_{L,2}=\left(\begin{array}{c}
d_{2}\\
-u_{2}\\
D_{1}
\end{array}
\right)_{L},\quad 
Q_{L,3}=\left(\begin{array}{c}
d_{3}\\
-u_{3}\\
D_{2}
\end{array}
\right)_{L}\sim (3,3^{\ast},0).
\end{equation}
Other two choices follow in an obvious way.
The right-handed quarks are placed in $SU(3)_L$ singlets. Their representation is unaffected by the representation choice for left-handed quarks, and is therefore the same for all three cases,
\begin{eqnarray}
&&  u_{R,i}\sim (3,1,\frac{2}{3}), \quad  U_{R}\sim (3,1,\frac{2}{3}),\\
&&  d_{R,i}\sim (3,1,-\frac{1}{3}), \quad  D_{R,1}\sim (3,1,-\frac{1}{3}), \quad  D_{R,2}\sim (3,1,-\frac{1}{3}),\quad i=1,2,3.
\end{eqnarray}
The fields $D_{1}$ and $D_{2}$ are new quarks with electric charge $-1/3$ and field $U$ new quark with electric charge $2/3$. These quarks mix with the SM quarks of the same electric charge. We will embrace this and study the mixing of quarks in its fullest. We will not introduce symmetries to forbid mixing between SM quarks and the exotic ones, like is often done in the literature.

\subsection{Scalar sector}
The minimal scalar sector that provides mass to all charged fermions at tree-level and breaks all the gauge symmetries is:
\begin{eqnarray}
&&\eta=\left(\begin{array}{c}
\eta^{+}\\
\eta^{0}\\
{\eta'}^{+}
\end{array}
\right)\sim (1,3,\frac{2}{3}),\quad  
\rho=\left(\begin{array}{c}
\rho^{0}\\
\rho^{-}\\
{\rho'}^{0}
\end{array}
\right),\quad \chi=\left(\begin{array}{c}
\chi^{0}\\
\chi^{-}\\
{\chi'}^{0}
\end{array}
\right)\sim (1,3,-\frac{1}{3}).\label{scalar triplets}
\end{eqnarray}
In principle all the neutral scalars can acquire a non-zero VEV. All the needed masses can be generated when each scalar triplet has one non-zero VEV\footnote{The inclusion of another non-zero $SU(3)_L$-breaking VEV would induce larger flavour violating effects than the current setup.},
\begin{eqnarray}\label{vacuum structure}
&&\langle\eta\rangle=\frac{1}{\sqrt{2}}\left(\begin{array}{c}
0\\
v_\eta\\
0
\end{array}
\right), \quad
\langle\rho\rangle=\frac{1}{\sqrt{2}}\left(\begin{array}{c}
v_\rho\\
0\\
0
\end{array}
\right),
\quad
\langle\chi\rangle=\frac{1}{\sqrt{2}}\left(\begin{array}{c}
0\\
0\\
v_\chi
\end{array}
\right).\label{vacuum}
\end{eqnarray}  
The breaking pattern, $SU(3)_L\times U(1)_X\to SU(2)_L\times U(1)_Y$, is induced by the VEV $v_\chi$ and  
the breaking of $SU(2)_L\times U(1)_Y$ into electromagnetism is 
caused by $v_\eta$ and $v_\rho$. We take $v_\eta$ and $v_\rho$ to be of the order of electroweak scale.  We also assume $v_\chi \gg v_\eta,v_\rho$.

Most general scalar potential is
\begin{align}\notag
& V=\mu^2_1 \eta^{\dagger}\eta+\mu_2^2 \rho^{\dagger}\rho+\mu_3^{2}\chi^{\dagger}\chi+(\mu^2_4 \rho^\dagger\chi+\text{h.c.})
+\lambda_1 (\eta^{\dagger}\eta)^2+\lambda_2 (\rho^{\dagger}\rho)^2
+\lambda_{3}(\chi^{\dagger}\chi)^2\\
& +\lambda_{12} (\eta^{\dagger}\eta)(\rho^{\dagger}\rho)+\lambda_{13} (\eta^{\dagger}\eta)(\chi^{\dagger}\chi)+\lambda_{23}(\rho^{\dagger}\rho)(\chi^{\dagger}\chi) + (\lambda_{1123} (\eta^{\dagger}\eta)(\rho^{\dagger}\chi)+\text{h.c.})\nonumber\\
& 
+(\lambda_{2223} (\rho^{\dagger}\rho)(\rho^{\dagger}\chi)+\text{h.c.})
+(\lambda_{3323} (\chi^{\dagger}\chi)(\rho^{\dagger}\chi)+\text{h.c.})
+(\lambda_{2323}(\rho^\dagger\chi)(\rho^\dagger\chi) + \text{h.c.})\\
& +\widetilde{\lambda}_{12} (\eta^{\dagger}\rho)(\rho^{\dagger}\eta)+\widetilde{\lambda}_{13} (\eta^{\dagger}\chi)(\chi^{\dagger}\eta)+\widetilde{\lambda}_{23}(\rho^{\dagger}\chi)(\chi^{\dagger}\rho) +(\widetilde{\lambda}_{1231} (\eta^{\dagger}\rho)(\chi^{\dagger}\eta)+\text{h.c.})\nonumber\\
&+\sqrt{2}f(\epsilon_{ijk}\eta^{i}\rho^{j}\chi^{k}+\text{h.c.}).\nonumber
\end{align}
The parameter $f$ has a mass dimension of one. We will assume $f\sim v_\chi$, in order to avoid the introduction of additional scales.

\subsection{Gauge sector}
The gauge sector of the 331-model contains 5 additional gauge bosons compared to the SM.  
The covariant derivative operating on a $SU(3)_L$-triplet is:
\begin{eqnarray}
&&D_{\mu}=\partial_{\mu}-ig_3 \sum_{a=1}^8 T_{a}W_{a\mu}-ig_{x}XB_{\mu},\nonumber
\end{eqnarray}
where  $g_3$ and $g_x$ are the $SU(3)_L$ and $U(1)_X$ gauge couplings, respectively. The $T_a=\lambda_a/2$ are the $SU(3)_L$ generators. 
The $SU(3)_L$ gauge bosons are: 
\begin{eqnarray}
 &&\sum_{a=1}^8 T_a W_{a\mu}=\frac{1}{\sqrt{2}}\left(\begin{array}{ccc}
\frac{1}{\sqrt{2}}W_{3\mu}+\frac{1}{\sqrt{6}}W_{8\mu} &
 {W}_{\mu}^{+} &  
{X}^{0}_{\mu}\\
 {W}^{-}_{\mu} &
-\frac{1}{\sqrt{2}}W_{3\mu}+\frac{1}{\sqrt{6}}W_{8\mu}& 
 {V}^{-}_{\mu}\\
 X^{0\ast}_{\mu} & 
 {V}^{+}_{\mu}&
-\frac{2}{\sqrt{6}}W_{8\mu}
\end{array}\right),\nonumber
\end{eqnarray} 
where we have denoted,
\begin{eqnarray}
{W}^{\pm}_\mu=\frac{1}{\sqrt{2}}(W_{1\mu}\mp iW_{2\mu}),\quad
{V}^{\mp}_\mu=\frac{1}{\sqrt{2}}(W_{6\mu}\mp iW_{7\mu}),\quad
{X}^{0}_\mu=\frac{1}{\sqrt{2}}(W_{4\mu}-iW_{5\mu}).\nonumber
\end{eqnarray}
The electrically neutral fields  $W_{3\mu}$, $W_{8\mu}$ and $B_\mu$ form photon, $Z$-boson and new heavy gauge boson $Z'$. The vacuum structure \eqref{vacuum structure} contains only one $SU(3)_L$-breaking VEV. Due to this the fields $W_{4\mu}$ and $W_{5\mu}$ do not mix with the other neutral gauge bosons. They have the same mass and are identified as a \emph{physical neutral non-hermitian gauge boson} $X^0_\mu\equiv\frac{1}{\sqrt{2}}(W_{4\mu}-iW_{5\mu})$. The off-diagonal gauge bosons ${W}^{\pm}_\mu$ and ${V}^{\pm}_\mu$ also don't mix, due to vacuum structure containing only one $SU(3)_L$-breaking VEV.  The $W^{\pm}_\mu$ is identified with the corresponding SM gauge boson and the $V^{\pm}_\mu$ as a new heavy charged gauge boson. The masses of the new gauge bosons are proportional to the $SU(3)_L\times U(1)_X$-breaking VEV $v_\chi$ and have their masses at TeV scale or higher.

The SM gauge boson masses are
\begin{equation}
m_W^2=  \frac{g_3^2}{4}({v_\eta}^2+v_\rho^2)\quad \textrm{and}\quad
m_Z^2 = \frac{g_3^2}{4\cos^2\theta_{W}}\left({v_\eta}^2+v_\rho^2\right)+\mathcal{O}\left(\delta^2\right),\nonumber
\end{equation}
with $\delta=v/v_\chi$, where $v=246$ GeV is the SM Higgs VEV.
The Weinberg angle is defined as
\begin{equation}
\cos^2 \theta_W=\frac{3g_3^2+g_x^2}{3g_3^2+4g_x^2}.
\end{equation}
At low energies the $g_3$ is identified with the SM $SU(2)_L$ gauge coupling.  The electroweak breaking VEVs $v_\eta$ and $v_\rho$ are related to the SM Higgs VEV through the relation
\begin{equation}\label{relation among vevs}
{v_\eta}^2+v_\rho^2=v^2,
\end{equation}
in order the $m_W$ and $m_Z$ to agree with the SM.
The heavy gauge boson masses are
\begin{equation}
m_V^2 = \frac{g_3^2}{4}(v_\chi^2+{v_\eta}^2),\quad
m_{Z'}^2 = \frac{g_x^2}{3}\frac{v_\chi^2}{\tan^2\theta_W}+\mathcal{O}\left(\delta^2\right),\quad
m_{X^0}^2 = \frac{g_3^2}{4}(v_\chi^2+v_\rho^2).
\end{equation}
The neutral states $Z$, $Z'$ and $X^0$ mediate quark FCNCs at tree-level.

\section{Scalar masses and eigenstates}\label{scalar masses}
We will pay special attention to neutral scalars and pseudo-scalars as they all mediate FCNCs. We are particularly interested in the $125$ GeV Higgs. It is potentially  dangerous  mediator of FCNCs due to its relative lightness. We will show that there is a natural suppression mechanism for FCNCs mediated by $125$ GeV Higgs. 
The neutral scalars are divided to real and imaginary parts as:
\begin{eqnarray}
&&\eta^{0}=\frac{1}{\sqrt{2}}(h_1+i \xi_1), \quad 
\rho^{0}=\frac{1}{\sqrt{2}}(h_2+i \xi_2), \quad {\rho'}^{0}=\frac{1}{\sqrt{2}}(h_5+i \xi_5),\\
&&\chi^{0}=\frac{1}{\sqrt{2}}(h_4+i \xi_4), \quad {\chi'}^{0}=\frac{1}{\sqrt{2}}(h_3+i \xi_3).\nonumber
\end{eqnarray}
We assume that all the parameters of the scalar potential are real and  therefore there is no mixing between real and imaginary parts of scalars. One of the five CP-even scalars is a would-be-Goldstone, giving mass to $X^0/X^{0\ast}$. Three of the five CP-odd scalars are would-be-Goldstones, providing the mass for $Z$, $Z'$ and $X^0/X^{0\ast}$. 

\subsection{CP-odd scalars}
The CP-odd scalar  mass term is,
\be\nonumber
 \mathcal{L}\supset \frac{1}{2}A^T M_\text{CP-odd}^2 A,\quad \textrm{where}\quad  A^T=(\xi_1,\xi_2,\xi_3,\xi_4,\xi_5)\quad \textrm{and},
\ee
\begin{eqnarray}
M_{cp-odd}^2=
\left(\begin{array}{ccccc}
f\frac{v_\rho v_\chi}{v_\eta} & f v_\chi & f v_\rho & 0 &  0\\
%%%
f v_\chi   & f\frac{v_\eta  v_\chi}{v_\rho} & f v_\eta & 0 & 0\\
%%%%
f v_\rho & f v_\eta & f\frac{v_\rho v_\eta}{v_\chi} &  0 & 0\\
%%%1
0 &
0 &
0 &
\left(\frac{\widetilde{\lambda}_{23}}{2}-\lambda_{2323}\right)v_\rho^2+f\frac{v_\rho v_\eta}{v_\chi} & \left(\frac{-\widetilde{\lambda}_{23}}{2}+\lambda_{2323}\right)v_\eta v_\chi -f v_\eta \\
%%%%
0 &
0 &
0 &
\left(\frac{-\widetilde{\lambda}_{23}}{2}+\lambda_{2323}\right)v_\eta v_\chi -f v_\eta &
\left(\frac{\widetilde{\lambda}_{23}}{2}-\lambda_{2323}\right)v_\chi^2+f\frac{v_\chi v_\eta}{v_\rho}
\end{array}\right).\nonumber
\end{eqnarray}
This matrix has three zero eigenvalues corresponding to the Goldstone bosons that give masses to $Z$, $Z'$ and the neutral non-Hermitian gauge boson $X^0_\mu$. The two non-zero eigenvalues correspond to physical CP-odd scalars $A_1$ and $A_2$.
The pseudo-scalar mass matrix is diagonalized as: 
\begin{equation}\label{cp odd diagonalization}
U^{A\dagger}M_\text{CP-odd}^2 U^{A}=M_{\textrm{A}}^2=\textrm{diag}(m_{A_1}^2,m_{A_2}^2,0,0,0) ,
\end{equation}
where
\begin{equation}\label{A1 and A2 mass}
m_{A_1}^2=fu\left(\frac{v_\rho}{v_\eta}+\frac{v_\eta}{v_\rho}+\frac{v_\rho v_\eta}{v_\chi^2}\right),\quad
m_{A_2}^2= (v_\chi^2+v_\rho^2)\left(\frac{\widetilde{\lambda}_{23}}{2}-\lambda_{2323}+\frac{f v_\eta}{v_\chi v_\rho}\right).
\end{equation}

The CP-odd mass eigenstates are defined as 
\begin{equation}\label{CP-odd rotation}
\left(\begin{array}{c}
A_1\\
A_2\\
G_1^0\\
G_2^0\\
G_3^0
\end{array}\right)
=U^{A\dagger} \left(\begin{array}{c}
\xi_1\\
\xi_2\\
\xi_3\\
\xi_4\\
\xi_5
\end{array}\right),
\end{equation}
where $G_i^0$, $i=1,2,3$, are massless Goldstones. The eigenvectors corresponding to $A_1$ and $A_2$ are\footnote{$M^2_{cp-odd}\bar{X}_{A_i}=m_{A_i}^2\bar{X}_{A_i}$, $i=1,2$.}
\begin{equation}\label{CP-odd eigenvectors}
\bar{X}_{A_1}=
\left(
\begin{array}{c}
U^{A}_{11}\\
U^{A}_{21}\\
U^{A}_{31}\\
U^{A}_{41}\\
U^{A}_{51}\\
\end{array}\right)=
\frac{1}{\sqrt{1+\frac{{v_\eta}^2 v_\rho^2}{v^2 v_\chi^2}}}\left(
\begin{array}{c}
\frac{v_\rho}{v}\\
\frac{v_\eta}{v}\\
\frac{v_\eta v_\rho}{v v_\chi}\\
0\\
0\\
\end{array}\right),\quad
\bar{X}_{A_2}=
\left(
\begin{array}{c}
U^{A}_{12}\\
U^{A}_{22}\\
U^{A}_{32}\\
U^{A}_{42}\\
U^{A}_{52}\\
\end{array}\right)=
\frac{1}{\sqrt{1+\frac{v_\rho^2}{v_\chi^2}}}\left(
\begin{array}{c}
0\\
0\\
0\\
-\frac{v_\rho}{v_\chi}\\
1
\end{array}\right).
\end{equation}
The pseudo-scalar $A_2$ will couple quite weakly to the SM quarks due to multiple zeros in the eigenvector\footnote{And in fact does not couple to charged leptons at tree-level at all!}. The structure of first two elements of $A_1$ is also important in cancellation of flavour violating effects, as we will see in next subsection.  

\subsection{CP-even scalars}
The CP-even scalar  mass term is,
\be\nonumber
 \mathcal{L}\supset \frac{1}{2}H^T M_{cp-even}^2 H,\quad \textrm{where}\quad  H^T=(h_1,h_2,h_3,h_4,h_5) \quad \textrm{and},
\ee
\be
\resizebox{1 \textwidth}{!} 
{
$ 
M_{cp-even}^2 =\label{cp-even mass matrix}\\
\left(\begin{array}{ccccc}
2\lambda_1{v_\eta}^2 +f\frac{v_\rho v_\chi}{v_\eta} & \lambda_{12}v_\eta v_\rho -f v_\chi & \lambda_{13} v_\eta v_\chi-f v_\rho & \lambda_{1123} v_\eta v_\rho & \lambda_{1123} v_\eta v_\chi \\
%%%
\lambda_{12}v_\eta v_\rho -f v_\chi   &2\lambda_2{v_\rho}^2 +f\frac{v_\eta  v_\chi}{v_\rho} & \lambda_{23} v_\rho v_\chi -fv_\eta & \lambda_{2223}  v_\rho^2 & \lambda_{2223} v_\rho v_\chi\\
%%%%
\lambda_{13}v_\eta v_\chi -f v_\rho & \lambda_{23}v_\rho v_\chi -f v_\eta  & 2\lambda_3{v_\chi}^2 +f\frac{v_\eta  v_\rho}{v_\chi} & \lambda_{3323} v_\rho v_\chi & \lambda_{3323}  v_\chi^2\\
%%%
\lambda_{1123} v_\rho v_\eta &
\lambda_{2223} v_\rho^2 &
\lambda_{3323} v_\rho v_\chi & 
\left(\frac{\widetilde{\lambda}_{23}}{2}+\lambda_{2323}\right)v_\rho^2 +f\frac{v_\eta  v_\rho}{v_\chi} & \left(\frac{\widetilde{\lambda}_{23}}{2}+\lambda_{2323}\right)v_\rho v_\chi +f v_\eta\\
%%%%
\lambda_{1123} v_\eta v_\chi &
\lambda_{2223} v_\rho v_\chi &
\lambda_{3323} v_\chi^2 &
\left(\frac{\widetilde{\lambda}_{23}}{2}+\lambda_{2323}\right)v_\rho v_\chi +f v_\eta &
\left(\frac{\widetilde{\lambda}_{23}}{2}+\lambda_{2323}\right)v_\chi^2 +f\frac{v_\eta v_\chi}{v_\rho} 
\end{array}\right).\nonumber
$
}
\ee
The matrix is diagonalized as:
\begin{equation}\label{cp even diagonalization}
U^{H\dagger}M_{CP-even}^2 U^{H}=M_{\textrm{H}}^2=\textrm{diag}(0,m_h^2,m_{H_1}^2,m_{H_2}^2,m_{H_3}^2).
\end{equation}
This matrix has one zero eigenvalue, corresponding to the Goldstone boson that gives mass to the  neutral non-Hermitian gauge boson $X^0$ and four non-zero eigenvalues corresponding to four physical CP-even scalars $h$, $H_1$, $H_2$ and $H_3$. One of the non-zero eigenvalues is $\mathcal{O}(v^2)$ and is identified with the $125$ GeV Higgs boson $h$ of the SM and therefore $m_h^2=(125 \textrm{ GeV})^2$.  
The three of the eigenvalues are $\mathcal{O}(v_\chi^2)$ and therefore very heavy. 

Assuming $v_\chi,f\gg v_\eta, v_\rho$, the upper left $2\times 2$-block in CP-even mass matrix are large compared to the surrounding elements. From this block one obtains the leading order contribution to the $125$ GeV Higgs and $H_1$ eigenvectors. The lower right $3\times 3$-block provides major contribution of the eigenvectors for $H_2, H_3$ and Goldstone $G^0_1$. 
The CP-even mass eigenstates are defined as:
\begin{equation}\label{CP-even rotation}
\left(\begin{array}{c}
G^0_1\\
h\\
H_1\\
H_2\\
H_3
\end{array}\right)
=U^{H\dagger} \left(\begin{array}{c}
h_1\\
h_2\\
h_3\\
h_4\\
h_5
\end{array}\right).
\end{equation}
The diagonalization matrix is
\begin{equation}\label{cp-even rotation matrix estimate}
U^H=\left(
\begin{array}{cccll}
0 &  \frac{v_\eta}{v_{}}+\mathcal{O}(\delta^2) & \frac{v_\rho}{v_{}}+\mathcal{O}(\delta^2) & \mathcal{O}(\delta) & \mathcal{O}(\delta^2)\\
%%%%%%%%%%%%%%%%%%%%%%%
0 & \frac{v_\rho}{v_{}}+\mathcal{O}(\delta^2) & -\frac{v_\eta}{v_{}}+\mathcal{O}(\delta^2) & \mathcal{O}(\delta) & \mathcal{O}(\delta^2)\\
%%%%%%%%%%%%%%%%%%%%%%%
0 & \mathcal{O}(\delta) & \mathcal{O}(\delta^2) & \mathcal{O}(1) & \mathcal{O}(1)\\
%%%%%%%%%%%%%%%%%%%%%%%
-\frac{1}{\sqrt{1+v_\rho^2/v_\chi^2}} & ~\mathcal{O}(\delta^2) & \mathcal{O}(\delta^3) & \mathcal{O}(\delta^2) & \mathcal{O}(\delta)\\
%%%%%%%%%%%%%%%%%%%%%%%
\frac{v_\rho/v_\chi}{\sqrt{1+v_\rho^2/v_\chi^2}} & \mathcal{O}(\delta) & \mathcal{O}(\delta^2) & \mathcal{O}(1) & \mathcal{O}(1)\\
\end{array}\right),
\end{equation}
where the columns correspond to the eigenvectors, according to Eq. \eqref{cp even diagonalization}. The first column is exact and corresponds to the Goldstone. Pay particular attention to the second column ($125$ GeV Higgs) and the third column ($H_1$). The leading order of first and second terms have simple expressions.
More accurate formula for these elements are,
\begin{eqnarray}
U_{12}^H &\approx & \frac{v_\eta}{v}-\frac{v_\rho^3 {v_\eta}^2({v_\eta}^2(2\lambda_1-\lambda_{12})+v_\rho^2(\lambda_{12}-2\lambda_2))}{f v_\chi v^5},\label{UH12}\\
U_{22}^H &\approx & \frac{v_\rho}{v}+\frac{v_\rho^2 {v_\eta}^3({v_\eta}^2(2\lambda_1-\lambda_{12})+v_\rho^2(\lambda_{12}-2\lambda_2))}{f v_\chi v^5}.\label{UH22}
\end{eqnarray}
For the $125$ GeV Higgs the cancellation between these terms causes major suppression for flavour violating couplings between SM quark, as we will see in Section \ref{neutral meson mixing}.
For the $H_1$ the cancellation takes place, not with itself, but with the $A_1$.
The mass of $H_1$ is
\be\label{H1 mass}
m_{H_1}^2\approx f v_\chi \left(\frac{v_\rho}{v_\eta}+\frac{v_\eta}{v_\rho}\right)
+ \frac{2v_\rho^2 {v_\eta}^2(\lambda_1+\lambda_2-\lambda_{12})}{v^2},
\ee
and by comparing to Eq. \eqref{A1 and A2 mass}, one can see that states $H_1$ and $A_1$ are almost mass degenerate. This is important for the cancellation of $H_1$ and $A_1$ contributions in neutral meson mixing. We return to this point later in Section \ref{neutral meson mixing}.

\section{Quark Yukawa couplings and masses}\label{quark yukawa couplings and masses}
In the SM only the Yukawa couplings after the diagonalization are known and determined by the fermion masses. The underlying Yukawa couplings before flavour rotation are unknown. In the SM the quark diagonalization matrices identically cancel for all neutral mediators when moving from gauge eigenstate to mass eigenstate. The only place where the quark rotation matrices do not cancel is the $W$ coupling to quarks. The resulting CKM matrix is the only observable giving information about the structure of the left-handed rotation matrices: $V_{\rm CKM}^\text{SM}=U_L^u U_L^{d\dagger}$. Therefore CKM is the only observable shedding some light to underlying Yukawa couplings, but the Yukawa couplings cannot still be determined from CKM. In the SM this is not a problem, as the fundamental Yukawa couplings don't appear anywhere after the flavour rotation and all the physical Yukawa couplings can be written in terms of the fermion masses. In 331-models this is, however, no longer the case: quark mass matrices are not diagonalized simultaneously with the Yukawa coupling matrices of multiple scalars. The quark rotation matrix elements become separately observable due to tree-level FCNCs.  
The structure of quark Yukawa couplings is essential in the attempts to predict the strength of the FCNCs, as the quark rotation matrices ultimately determine the strength of the flavour violating couplings.  Some assumptions of the underlying quark Yukawa structure has to be made in order to proceed. Before going into the hierarchies, we present the general structure of Yukawa couplings for all the scalars and mass matrices, for each generation assignment.

\subsection{Up-type quark Yukawa couplings and masses\label{sec:up}}
All the scalar triplets couple to the quarks. 
The up-type Yukawa couplings for the case where the generation $\mathcal{G}$ is in triplet are : 
\begin{eqnarray}
\mathcal{L}_\text{up}
&=&\sum_{\alpha\neq \mathcal{G}}\sum_{\gamma=1}^{4}(y^u_{\eta^{\ast}})_{\alpha\gamma}\bar{Q}'_{L,\alpha}\eta^{\ast} ~u'_{R,\gamma}
+\sum_{\gamma=1}^{4}(y^u_\rho)_{\mathcal{G}\gamma}\bar{Q}'_{L,\mathcal{G}}\rho ~u'_{R,\gamma}
+\sum_{\gamma=1}^{4}(y^u_\chi)_{\mathcal{G}\gamma}\bar{Q}'_{L,\mathcal{G}}\chi ~u'_{R,\gamma}+\hc,\nonumber\label{u-yukawa interaction}
\end{eqnarray}
where $u'_{R}=({u'}_{R,1},{u'}_{R,2},{u'}_{R,3},{U'}_{R})$ and the $\alpha$ labels the two generations that are in antitriplet.
 The Yukawa couplings of up-type quarks to scalars and pseudo-scalars before flavour rotation are
\begin{eqnarray}\label{up yukawas prime}
\mathcal{L}_{\textrm{up}}&=&
\sum_{\phi}\frac{1}{\sqrt{2}}\bar{u}'_{L}({\Gamma'}^u_{\phi, \mathcal{G}})u'_{R}~\phi
+ i\sum_{i=1,2}\frac{1}{\sqrt{2}}\bar{u}'_{L}({\Gamma'}^u_{A_i, \mathcal{G}})u'_{R}~A_i +h.c.\label{up yukawas prime}
\end{eqnarray}
where the primes denote  gauge eigenstates  and  the coupling matrices for ${\Gamma'}^u_{\phi,\text{1st}}$, ${\Gamma'}^u_{\phi,\text{2nd}}$, ${\Gamma'}^u_{\phi,\text{3rd}}$, ${\Gamma'}^u_{A_i,\text{1st}}$, ${\Gamma'}^u_{A_i,\text{2nd}}$, ${\Gamma'}^u_{A_i,\text{3rd}}$ are:
\begin{equation}
%\resizebox{1.0 \textwidth}{!} 
%{
%$ 
\left(
\begin{array}{cccc}
U^{H}_{2\phi}(y^{u}_{\rho})_{1\gamma}+U^{H}_{4\phi}(y^{u}_{\chi})_{1\gamma}\\
-U^{H}_{1\phi}(y^{u}_{\eta^*})_{2\gamma}\\
-U^{H}_{1\phi}(y^{u}_{\eta^*})_{3\gamma}\\
U^{H}_{5\phi}(y^{u}_{\rho})_{1\gamma}+U^{H}_{3\phi}(y^{u}_{\chi})_{1\gamma} \nonumber
\end{array}
\right),\quad
\left(
\begin{array}{cccc}
-U^{H}_{1\phi}(y^{u}_{\eta^*})_{1\gamma}\\
U^{H}_{2\phi}(y^{u}_{\rho})_{2\gamma}+U^{H}_{4\phi}(y^{u}_{\chi})_{2\gamma}\\
-U^{H}_{1\phi}(y^{u}_{\eta^*})_{3\gamma}\\
U^{H}_{5\phi}(y^{u}_{\rho})_{2\gamma}+U^{H}_{\phi}(y^{u}_{\chi})_{2\gamma} \nonumber
\end{array}
\right),\quad
\left(
\begin{array}{cccc}
-U^{H}_{1\phi}(y^{u}_{\eta^*})_{1\gamma}\\
-U^{H}_{1\phi}(y^{u}_{\eta^*})_{2\gamma}\\
U^{H}_{2\phi}(y^{u}_{\rho})_{3\gamma}+U^{H}_{4\phi}(y^{u}_{\chi})_{3\gamma}\\
U^{H}_{5\phi}(y^{u}_{\rho})_{3\gamma}+U^{H}_{3\phi}(y^{u}_{\chi})_{3\gamma} \nonumber
\end{array}
\right),
%$
%}
\end{equation}
\begin{equation}
%\resizebox{1.0 \textwidth}{!} 
%{
%$ 
\left(
\begin{array}{cccc}
U^{A}_{2i}(y^{u}_{\rho})_{1\gamma}+U^{A}_{4i}(y^{u}_{\chi})_{1\gamma}\\
U^{A}_{1i}(y^{u}_{\eta^*})_{2\gamma}\\
U^{A}_{1i}(y^{u}_{\eta^*})_{3\gamma}\\
U^{A}_{5i}(y^{u}_{\rho})_{1\gamma}+U^{A}_{3i}(y^{u}_{\chi})_{1\gamma} \nonumber
\end{array}
\right),\quad
\left(
\begin{array}{cccc}
U^{A}_{1i}(y^{u}_{\eta^*})_{1\gamma}\\
U^{A}_{2i}(y^{u}_{\rho})_{2\gamma}+U^{A}_{4i}(y^{u}_{\chi})_{2\gamma}\\
U^{A}_{1i}(y^{u}_{\eta^*})_{3\gamma}\\
U^{A}_{5i}(y^{u}_{\rho})_{2\gamma}+U^{A}_{3i}(y^{u}_{\chi})_{2\gamma} \nonumber
\end{array}
\right),\quad
\left(
\begin{array}{cccc}
U^{A}_{1i}(y^{u}_{\eta^*})_{1\gamma}\\
U^{A}_{1i}(y^{u}_{\eta^*})_{2\gamma}\\
U^{A}_{2i}(y^{u}_{\rho})_{3\gamma}+U^{A}_{4i}(y^{u}_{\chi})_{3\gamma}\\
U^{A}_{5i}(y^{u}_{\rho})_{3\gamma}+U^{A}_{3i}(y^{u}_{\chi})_{3\gamma} \nonumber
\end{array}
\right),
%$
%}
\end{equation}
respectively and $U^H$ and $U^A$ are defined in Eqs. \eqref{cp odd diagonalization} and \eqref{cp even diagonalization}. The $\phi=2,3,4,5$ corresponds to $h, H_1, H_2, H_3$, respectively. The $\gamma$ runs from $1$ to $4$ to give $4\times 4$ matrix.

The up-type quark masses are generated by the terms in the Eq. (\ref{u-yukawa interaction}).
The up-quark mass matrix in the basis,
\begin{equation}
\mathcal{L}_\text{up-mass}=\bar{u}'_{L} m^u_{\mathcal{G}} u'_R+h.c.,
\end{equation}
where $m^u_\text{1st}$, $m^u_\text{2nd}$, $m^u_\text{3rd}$ are,
\begin{eqnarray}
\frac{1}{\sqrt{2}}\left(
\begin{array}{c}
 v_\rho (y^{u}_{\rho})_{1\gamma}\\
-v_\eta (y^{u}_{\eta^*})_{2\gamma}\\
-v_\eta (y^{u}_{\eta^*})_{3\gamma}\\
v_\chi(y^{u}_{\chi})_{1\gamma}\\
\end{array}
\right),\quad
\frac{1}{\sqrt{2}}\left(
\begin{array}{c}
-v_\eta (y^{u}_{\eta^*})_{1\gamma}\\
 v_\rho (y^{u}_{\rho})_{2\gamma}\\
-v_\eta (y^{u}_{\eta^*})_{3\gamma}\\
v_\chi(y^{u}_{\chi})_{2\gamma}\\
\end{array}
\right),\quad
\frac{1}{\sqrt{2}}\left(
\begin{array}{c}
-v_\eta (y^{u}_{\eta^*})_{1\gamma}\\
-v_\eta (y^{u}_{\eta^*})_{2\gamma}\\
 v_\rho (y^{u}_{\rho})_{3\gamma}\\
v_\chi(y^{u}_{\chi})_{3\gamma}\\
\end{array}
\right),\label{up generation assignment mass matrices}
\end{eqnarray}
respectively. We have written the mass matrix in column form for the sake of brevity.

\subsection{Down-type quark Yukawa couplings and masses\label{sec:down}}
 The down-type quark Yukawa couplings are  written  similarly to the up-type couplings. 
For the case where generation $\mathcal{G}$ is in triplet are, 
\begin{eqnarray}
\mathcal{L}_\text{down}
&=&
\sum_{\alpha\neq \mathcal{G}}\sum_{\gamma=1}^{5}(y^d_{\rho^{\ast}})_{\alpha\gamma}\bar{Q}'_{L,\alpha}\rho^{\ast} d'_{R,\gamma}
+\sum_{\alpha\neq \mathcal{G}}\sum_{\gamma=1}^{5}(y^d_{\chi^{\ast}})_{\alpha\gamma}\bar{Q}'_{L,\alpha}\chi^{\ast} d'_{R,\gamma}
+\sum_{\gamma=1}^{5}(y^d_{\eta})_{\mathcal{G}\gamma}\bar{Q}'_{L,\mathcal{G}}\eta d'_{R,\gamma}
\nonumber\\
&+& h.c.,\label{d-yukawa interaction}
\end{eqnarray}
where $d'_{R}=({d'}_{R,1},{d'}_{R,2},{d'}_{R,3},{D'}_{R,1},{D'}_{R,2})$. The $\alpha$ labels the quark generations that are in antitriplet. The Yukawa couplings before flavour rotation are,
\begin{eqnarray}
\mathcal{L}_{\textrm{down}}&=&
\sum_{\phi}\frac{1}{\sqrt{2}}\bar{d}'_{L}({\Gamma'}^d_{\phi, \mathcal{G}})d'_{R}~\phi
+ i\sum_{i=1,2}\frac{1}{\sqrt{2}}\bar{d}'_{L}({\Gamma'}^d_{A_i, \mathcal{G}})d'_{R}~A_i +h.c.,\label{down yukawas prime}
\end{eqnarray}
where the primes denote  gauge eigenstates  and  the coupling matrices for ${\Gamma'}^d_{\phi,\text{1st}}$, ${\Gamma'}^d_{\phi,\text{2nd}}$, ${\Gamma'}^d_{\phi,\text{3rd}}$, ${\Gamma'}^d_{A_i,\text{1st}}$, ${\Gamma'}^d_{A_i,\text{2nd}}$, ${\Gamma'}^d_{A_i,\text{3rd}}$ are:

\begin{equation}
%\resizebox{1.0 \textwidth}{!} 
%{
%$ 
\left(
\begin{array}{cccc}
U^{H}_{1\phi}(y^{d}_{\eta})_{1\gamma} \\
U^{H}_{2\phi}(y^{d}_{\rho^*})_{2\gamma}+U^{H}_{4\phi}(y^{d}_{\chi^*})_{2\gamma}  \\
U^{H}_{2\phi}(y^{d}_{\rho^*})_{3\gamma}+U^{H}_{4\phi}(y^{d}_{\chi^*})_{3\gamma}  \\
U^{H}_{5\phi}(y^{d}_{\rho^*})_{2\gamma}+U^{H}_{3\phi}(y^{d}_{\chi^*})_{2\gamma} \\
U^{H}_{5\phi}(y^{d}_{\rho^*})_{3\gamma}+U^{H}_{3\phi}(y^{d}_{\chi^*})_{3\gamma}  \nonumber
\end{array}
\right),\quad
\left(
\begin{array}{cccc}
U^{H}_{2\phi}(y^{d}_{\rho^*})_{1\gamma}+U^{H}_{\phi}(y^{d}_{\chi^*})_{1\gamma}  \\
U^{H}_{1\phi}(y^{d}_{\eta})_{2\gamma} \\
U^{H}_{2\phi}(y^{d}_{\rho^*})_{3\gamma}+U^{H}_{4\phi}(y^{d}_{\chi^*})_{3\gamma}  \\
U^{H}_{5\phi}(y^{d}_{\rho^*})_{1\gamma}+U^{H}_{3\phi}(y^{d}_{\chi^*})_{1\gamma} \\
U^{H}_{5\phi}(y^{d}_{\rho^*})_{3\gamma}+U^{H}_{3\phi}(y^{d}_{\chi^*})_{3\gamma} \nonumber
\end{array}
\right),\quad
\left(
\begin{array}{cccc}
U^{H}_{2\phi}(y^{d}_{\rho^*})_{1\gamma}+U^{H}_{4\phi}(y^{d}_{\chi^*})_{1\gamma}  \\
U^{H}_{2\phi}(y^{d}_{\rho^*})_{2\gamma}+U^{H}_{4\phi}(y^{d}_{\chi^*})_{2\gamma}  \\
U^{H}_{1\phi}(y^{d}_{\eta})_{3\gamma} \\
U^{H}_{5\phi}(y^{d}_{\rho^*})_{1\gamma}+U^{H}_{3\phi}(y^{d}_{\chi^*})_{1\gamma} \\
U^{H}_{5\phi}(y^{d}_{\rho^*})_{2\gamma}+U^{H}_{3\phi}(y^{d}_{\chi^*})_{2\gamma}  \nonumber
\end{array}
\right),
%$
%}
\end{equation}
\begin{equation}
%\resizebox{1.0 \textwidth}{!} 
%{
%$ 
\left(
\begin{array}{cccc}
U^{A}_{1i}(y^{d}_{\eta})_{1\gamma} \\
-U^{A}_{2i}(y^{d}_{\rho^*})_{2\gamma}-U^{A}_{4i}(y^{d}_{\chi^*})_{2\gamma}  \\
-U^{A}_{2i}(y^{d}_{\rho^*})_{3\gamma}-U^{A}_{4i}(y^{d}_{\chi^*})_{3\gamma}  \\
-U^{A}_{5i}(y^{d}_{\rho^*})_{2\gamma}-U^{A}_{3i}(y^{d}_{\chi^*})_{2\gamma} \\
-U^{A}_{5i}(y^{d}_{\rho^*})_{3\gamma}-U^{A}_{3i}(y^{d}_{\chi^*})_{3\gamma}  \nonumber
\end{array}
\right),\quad
\left(
\begin{array}{cccc}
-U^{A}_{2i}(y^{d}_{\rho^*})_{1\gamma}-U^{A}_{4i}(y^{d}_{\chi^*})_{1\gamma}  \\
U^{A}_{1i}(y^{d}_{\eta})_{2\gamma} \\
-U^{A}_{2i}(y^{d}_{\rho^*})_{3\gamma}-U^{A}_{4i}(y^{d}_{\chi^*})_{3\gamma}  \\
-U^{A}_{5i}(y^{d}_{\rho^*})_{1\gamma}-U^{A}_{3i}(y^{d}_{\chi^*})_{1\gamma} \\
-U^{A}_{5i}(y^{d}_{\rho^*})_{3\gamma}-U^{A}_{3i}(y^{d}_{\chi^*})_{3\gamma} \nonumber
\end{array}
\right),\quad
\left(
\begin{array}{cccc}
-U^{A}_{2i}(y^{d}_{\rho^*})_{1\gamma}-U^{A}_{4i}(y^{d}_{\chi^*})_{1\gamma}  \\
-U^{A}_{2i}(y^{d}_{\rho^*})_{2\gamma}-U^{A}_{4i}(y^{d}_{\chi^*})_{2\gamma}  \\
U^{A}_{1i}(y^{d}_{\eta})_{3\gamma} \\
-U^{A}_{5i}(y^{d}_{\rho^*})_{1\gamma}-U^{A}_{3i}(y^{d}_{\chi^*})_{1\gamma} \\
-U^{A}_{5i}(y^{d}_{\rho^*})_{2\gamma}-U^{A}_{3i}(y^{d}_{\chi^*})_{2\gamma}  \nonumber
\end{array}
\right),
%$
%}
\end{equation}
respectively. The $\phi=2,3,4,5$ corresponds to $h, H_1, H_2, H_3$, respectively. The $\gamma$ runs from $1$ to $5$ to give $5\times 5$ matrix.

The down-type quark masses are generated by the terms in  Eq. (\ref{d-yukawa interaction}).
The down-quark mass matrix in the basis:
\begin{equation}
\mathcal{L}_\text{down-mass}=\bar{d}'_{L} m^d_{\mathcal{G}} d'_R+h.c.,
\end{equation}
where $m^d_\text{1st}$, $m^d_\text{2nd}$, $m^d_\text{3rd}$,
\begin{eqnarray}
\frac{1}{\sqrt{2}}\left(
\begin{array}{c}
v_\eta (y^{d}_{\eta})_{1\gamma} \\
v_\rho  (y^{d}_{\rho^*})_{2\gamma}  \\
v_\rho (y^{d}_{\rho^*})_{3\gamma}  \\
v_\chi(y^{d}_{\chi^*})_{2\gamma} \\
v_\chi(y^{d}_{\chi^*})_{3\gamma}  \\
\end{array}
\right),\quad
\frac{1}{\sqrt{2}}\left(
\begin{array}{c}
v_\rho  (y^{d}_{\rho^*})_{1\gamma}  \\
v_\eta (y^{d}_{\eta})_{2\gamma} \\
v_\rho (y^{d}_{\rho^*})_{3\gamma}  \\
v_\chi(y^{d}_{\chi^*})_{1\gamma} \\
v_\chi(y^{d}_{\chi^*})_{3\gamma}  \\
\end{array}
\right),\quad
\frac{1}{\sqrt{2}}\left(
\begin{array}{c}
v_\rho  (y^{d}_{\rho^*})_{1\gamma}  \\
v_\rho (y^{d}_{\rho^*})_{2\gamma}  \\
v_\eta (y^{d}_{\eta})_{3\gamma} \\
v_\chi(y^{d}_{\chi^*})_{1\gamma} \\
v_\chi(y^{d}_{\chi^*})_{2\gamma}  \\
\end{array}
\right),\label{down generation assignment mass matrices}
\end{eqnarray}
respectively. 
The quark mass matrices, $m^u$ and $m^d$ are diagonalized through biunitary transformation:
\be\label{quark mass matrix diagonalization}
m^u_{\textrm{diag}}=U_L^u m^u U_R^{u\dagger}, \quad \textrm{and}\quad
m^d_{\textrm{diag}}=U_L^d m^d U_R^{d\dagger}.
\ee

\subsection{CKM naturality}\label{CKM naturality}

The constraints on the quark Yukawa couplings come from SM quark masses and the elements of the CKM matrix. We assume that Yukawa couplings have hierarchical structure, such that the quark masses are generated without miraculous cancellation between parameters, while remaining agnostic about the origin of this hierarchy. 
The second assumption we make is that the CKM matrix is produced without significant cancellations between up- and down-quark rotation matrix elements. 

Due to exotic quarks the CKM matrix is not a $3\times 3$ matrix, but a $4\times 5$ matrix,
\be
V_\text{CKM}^{331}=U_L^u\left(\begin{array}{ccccc}
1 & 0 & 0 & 0 & 0\\
0 & 1 & 0 & 0 & 0\\
0 & 0 & 1 & 0 & 0\\
0 & 0 & 0 & 0 & 0
\end{array}\right)
U_L^{d\dagger},
\ee
with
\begin{eqnarray}
\mathcal{L}_\text{CKM}&&=\frac{g_{3}}{\sqrt{2}}\bar u_L\gamma^{\mu}
V_\text{CKM}^{331}
d_L{W}^{+}_{\mu}
+h.c.,\label{charged gauge boson couplings}
\end{eqnarray}
and $u_L = ({u}_{1}~{u}_{2}~{u}_{3}~{U})_L^T$, $d_L = ({d}_{1}~{d}_{2}~{d}_{3}~{D}_{1}~D_{2})_L^T$. The upper-left $3\times 3$ - block of $V_\text{CKM}^{331}$ corresponds to the SM CKM matrix. 
The CKM matrix exhibits distinct hierarchy \cite{ParticleDataGroup:2022pth},
\be
|V_{\rm CKM}^{\rm SM}| = \left(\begin{array}{ccc}
0.97420\pm 0.00021 & 0.2243\pm 0.0005 & (3.94\pm 0.36)\times 10^{-3}\\
0.218\pm 0.004 & 0.997\pm 0.017 & (42.2\pm 0.8)\times 10^{-3}\\
(8.1\pm 0.5)\times 10^{-3} & (39.4\pm 2.3)\times 10^{-3} & 1.019\pm 0.025
\end{array}\right).
\ee
The magnitude of the CKM matrix elements are roughly powers of the sine of Cabibbo angle $\epsilon\simeq 0.23$,
\be\label{SM CKM hierarchy}
V_{\rm CKM}^{\rm SM}\sim\left(\begin{array}{ccc}
1 & \epsilon & \epsilon^3\\
\epsilon & 1 & \epsilon^2\\
\epsilon^3 & \epsilon^2 & 1
\end{array}\right).
\ee
This structure is obtained in upper $3\times 3$ block of $V_\text{CKM}^{331}$, without miraculous cancellations, when the $U_L^u$ and $U_L^d$ have the same texture in their upper $3\times 3$ block corresponding to SM quarks. 
This is the approach we adopt here. We shall assume that the Yukawa couplings are such that this is naturally realized. We dub this \emph{CKM-naturality}.
This is achieved by having preset hierarchy in the Yukawa couplings. The hierarchy of rows sets the structure of the left-handed rotation matrices and hierarchy of columns sets the structure of the right-handed rotation matrices. We write the Yukawa couplings in the following manner to emphasize the effect of the individual Yukawa couplings on the structure of rotation matrices,
\be\label{yukawa structure}
(y^q)_{ij} = c^q_{ij}\epsilon^{L^q_i+R^q_j},
\ee
where $c^q_{ij}$ is an order one coupling and $i,j$ run from $1$ to $4$ for up-type quarks and from $1$ to $5$ for down. The powers ${L}^q_i$ and $R^q_j$ determine the structure of the rotation matrices: the values of ${L}^q_i$ determine the left-handed rotation matrix and $R^q_j$ that of the right-handed. 
This is reminiscent of notation used in the context of the Froggatt-Nielsen mechanism \cite{Froggatt:1978nt}. Difference is that now, the powers ${L}^q_i$ and $R^q_j$ are not related to any underlying flavour symmetry, like in the FN models. We will use this as a book keeping device and to make the hierarchical structure of Yukawa couplings more transparent. The powers of $\epsilon$ are larger for lighter generations, making the corresponding entries smaller, and vice versa. 

Let us write the up and down quark mass matrices in a form that is the same for all generation assignments (we have suppressed for brevity the scalar labels present in the previous subsections \ref{sec:up} and \ref{sec:down}):
\be\label{up mass matrix}
m^u\equiv\frac{1}{\sqrt{2}}\left(
\begin{array}{cccc}
v_\eta (y^{u})_{1\gamma}\\
v_\eta (y^{u})_{2\gamma}\\
v_\eta (y^{u})_{3\gamma}\\
v_\chi (y^{u})_{4\gamma}
\end{array}
\right)
=
\frac{v_\eta}{\sqrt{2}}\left(
\begin{array}{l}
(c^{u})_{1\gamma}\epsilon^{\widetilde{L}^u_1+R^u_{\gamma}}\\
(c^{u})_{2\gamma}\epsilon^{\widetilde{L}^u_2+R^u_{\gamma}}\\
(c^{u})_{3\gamma}\epsilon^{\widetilde{L}^u_3+R^u_{\gamma}}\\
(c^{u})_{4\gamma}\epsilon^{\widetilde{L}^u_4+R^u_{\gamma}}
\end{array}
\right),
\ee
\be\label{down mass matrix}
m^d\equiv\frac{1}{\sqrt{2}}\left(
\begin{array}{l}
v_\eta (y^{d})_{1\gamma}\\
v_\eta (y^{d})_{2\gamma}\\
v_\eta (y^{d})_{3\gamma}\\
v_\chi (y^{d})_{4\gamma}\\
v_\chi (y^{d})_{5\gamma}
\end{array}
\right)
=
\frac{v_\eta}{\sqrt{2}}\left(
\begin{array}{l}
(c^{d})_{1\gamma}\epsilon^{\widetilde{L}^d_1+R^d_{\gamma}}\\
(c^{d})_{2\gamma}\epsilon^{\widetilde{L}^d_2+R^d_{\gamma}}\\
(c^{d})_{3\gamma}\epsilon^{\widetilde{L}^d_3+R^d_{\gamma}}\\
(c^{d})_{4\gamma}\epsilon^{\widetilde{L}^d_4+R^d_{\gamma}}\\
(c^{d})_{5\gamma}\epsilon^{\widetilde{L}^d_5+R^d_{\gamma}}
\end{array}
\right),
\ee
where we have defined $\widetilde{L}^q_i\equiv L^q_i$, for $i=1,2,3$ and absorbed the $SU(3)_L$-breaking VEV into the left-handed powers of exotic quarks,
\be
\widetilde{L}^u_4 = L_U+\frac{\log(v_\chi/v_\eta)}{\log(\epsilon)},\quad 
\widetilde{L}^d_4 = L_{D_1}+\frac{\log(v_\chi/v_\eta)}{\log(\epsilon)},\quad
\widetilde{L}^d_5 = L_{D_2}+\frac{\log(v_\chi/v_\eta)}{\log(\epsilon)},
\ee
with ${L}^u_4\equiv L_U$, ${L}^d_4\equiv L_{D_1}$ and ${L}^d_5\equiv L_{D_2}$.
The order-one couplings $c^q_{ij}$ are different for each generation assignment, and obtained by comparing mass matrices in the Eqs. \eqref{up generation assignment mass matrices} and \eqref{down generation assignment mass matrices}  to the above. The mass matrix elements have equal magnitude for all generation assignment.
As long as $\widetilde{L}^q_1\geq \widetilde{L}^q_2\dots$ and $R^q_1\geq R^q_2\dots$, the rotation matrix elements will have the following order of magnitude in rotation matrix elements,
\be
(U^q_L)_{ij}\sim \epsilon^{|\widetilde{L}^q_i-\widetilde{L}^q_j|}\quad \textrm{and} \quad
(U^q_R)_{ij}\sim \epsilon^{|R^q_i-R^q_j|},
\ee
and masses $m^q_i\sim v_\eta\epsilon^{\widetilde{L}^q_i+R^q_i}$ \cite{Froggatt:1978nt}. We assume the following hierarchies in the quark mass matrices: $m^u_{1j}\leq m^u_{2j}\leq m^u_{3j}\leq m^u_{4j}$ and $m^d_{1j}\leq m^d_{2j}\leq m^d_{3j}\leq m^d_{4j}\leq m^d_{5j}$. This will generate CKM-like hierarchy when left-handed powers are properly chosen. We also assume the hierarchy  $m^u_{i1}\leq m^u_{i2}\leq m^u_{i3}\leq m^u_{i4}$ and $m^d_{i1}\leq m^d_{i2}\leq m^d_{i3}\leq m^d_{i4}\leq m^d_{i5}$, which will affect the right-handed rotation matrix. The right-handed rotation matrix is, however, not relevant here. 

We choose $L_1\equiv L^u_1=L^d_1$, $L_2\equiv L^u_2=L^d_2$ and $L_3\equiv L^u_3=L^d_3$. The $L_1$, $L_2$ and $L_3$ are fixed by the CKM-naturality, up to an additive constant. The CKM-structure of \eqref{SM CKM hierarchy} is obtained by choosing $L_1=3$, $L_2=2$ and $L_3=0$.
The $L_U$, $L_{D_1}$, $L_{D_2}$ are free parameters, whose main function is in determining the exotic elements of the left-handed rotation matrices. The exotic elements of the quark rotation matrices receive additional suppression from large difference between electroweak scale and the $SU(3)_L$-breaking scale. 
The logarithmic terms act as effective left-handed "charges", if one uses analogy to the FN-mechanism. 
The mass matrices exhibit hierarchy between exotic quarks and SM quarks, due to $v_\chi\gg v_\eta,v_\rho$. We assume that the exotic rows in the quark mass matrices are larger than those in the third row in the same columns. This is a necessary requirement as otherwise the exotic quarks would be too light. This restricts the $L_U$, $L_{D_1}$ and $L_{D_2}$ not to be too large.    
The freedom in choosing right-handed charges is used to give correct masses for SM quarks.  

The order of magnitude of left-handed rotation matrices is
\be\label{up rotation texture}
U_L^u\sim\left(\begin{array}{cccc}
1 & \epsilon & \epsilon^3 & \delta\epsilon^{3-L_U}\\
\epsilon & 1 & \epsilon^2 & \delta\epsilon^{2-L_U}\\
\epsilon^3 & \epsilon^2 & 1 & \delta\epsilon^{-L_U}\\
\delta\epsilon^{3-L_U} & \delta\epsilon^{2-L_U} & \delta\epsilon^{-L_U} & 1
\end{array}\right),
\ee
\be\label{down rotation texture}
U_L^d\sim\left(\begin{array}{ccccc}
1 & \epsilon & \epsilon^3 & \delta\epsilon^{3-L_{D_1}} & \delta\epsilon^{3-L_{D_2}}\\
\epsilon & 1 & \epsilon^2 & \delta\epsilon^{2-L_{D_1}} & \delta\epsilon^{2-L_{D_2}}\\
\epsilon^3 & \epsilon^2 & 1 & \delta\epsilon^{-L_{D_1}} & \delta\epsilon^{-L_{D_2}}\\
\delta\epsilon^{3-L_{D_1}} & \delta\epsilon^{2-L_{D_1}} & \delta\epsilon^{-L_{D_1}} & 1 & \epsilon^{L_{D_1}-L_{D_2}}\\
\delta\epsilon^{3-L_{D_2}} & \delta\epsilon^{2-L_{D_2}} & \delta\epsilon^{-L_{D_2}} & \delta\epsilon^{L_{D_1}-L_{D_2}} & 1
\end{array}\right),
\ee
which are independent of the right-handed powers. 
One notices that the mixing of exotic quarks with the SM quarks is suppressed by $\delta$. This suppression is due to exotic quark mass terms being proportional to $SU(3)_L$-breaking VEV, $v_\chi$. 
With the above rotation textures the CKM matrix texture is:
\be\label{331 ckm}
V_{\rm CKM}^{331}\sim\left(\begin{array}{ccccc}
1 & 
\epsilon & 
\epsilon^3 &
\delta\epsilon^{3-L_{D_1}} &
\delta\epsilon^{3-L_{D_2}}\\
%%%%%%%%%%%%%%%%%%%%%%%%%%
\epsilon &
1 &
\epsilon^2 &
\delta\epsilon^{2-L_{D_1}} &
\delta\epsilon^{2-L_{D_2}}\\
%%%%%%%%%%%%%%%%%%%%%%%%%%
\epsilon^3 &
\epsilon^2 &
1 &
\delta\epsilon^{-L_{D_1}} &
\delta\epsilon^{-L_{D_2}}\\
%%%%%%%%%%%%%%%%%%%%%%%%%%
\delta\epsilon^{3-L_{U}} &
\delta\epsilon^{2-L_{U}} &
\delta\epsilon^{-L_{U}} & 
\delta^2\epsilon^{-L_{D_1}-L_U} &
\delta^2\epsilon^{-L_{D_2}-L_U}
\end{array}\right).
\ee
The upper-left $3\times 3$ block corresponds to SM CKM matrix. The other element involving mixing with exotic quarks are suppressed by $SU(3)_L$-breaking scale. 

The outcome of the CKM naturality is that 
%the small rotation matrix elements are evenly distributed.
the off-diagonal elements of the quark rotation matrices are suppressed.
If the left-handed quark rotation matrices were to have democratic texture, that is, elements of order one throughout the matrix, there would have to be accidental cancellations between elements of $U_L^u$ and $U_L^{d\dagger}$ in order to produce CKM matrix. We deem this unnatural. Democratic rotation matrices would also lead into large flavour violating effects. %, as we will see shortly. 
The CKM naturality (rotation matrices have CKM texture) helps to minimize the magnitude of rotation matrix elements, and therefore the flavour violating effects as well.

\section{Neutral meson mixing and mediator couplings}\label{neutral meson mixing}

The neutral mesons composed of quarks with different flavour form two eigenstates dubbed long-lived and short-lived. These states have slightly different masses. 
The current experimental measurements for neutral meson mass differences are consistent with the prediction from the SM\footnote{The SM computation of neutral meson mass difference has large uncertainties due to QCD corrections \cite{Wang:2019try, Wang:2022lfq, DeBruyn:2022zhw}}.
In the SM the leading contribution to operators contributing to the neutral meson mass difference is generated at 1-loop level by $W^\pm$-box diagrams. In 331-model the BSM contribution is however generated already at the tree-level. This imposes stringent bounds on the model parameters.  
We take into account all possible mediators of neutral meson mixing and analyze in detail the magnitude of their contribution to neutral meson mass difference in order to shed some light to inner workings of the flavour violating effects that take place in 331-model of this type.     

We follow \cite{Gabbiani:1996hi} for the computation of neutral meson mass difference. The effective Hamiltonian for neutral kaon mixing at tree-level is
\begin{equation}
\mathcal{H}_{eff}^{K^0\textrm{-}\bar{K}^0}= 
C_1^{ds} (\bar s_L \gamma^\mu d_L)^2
+C^{ds}_2 (\bar{s}_R d_L)^2 
+ \widetilde{C}^{ds}_2 (\bar{s}_L d_R)^2 
+ C^{ds}_4 (\bar{s}_R d_L)(\bar{s}_L d_R),
\end{equation}
where the Wilson coefficients are,
\begin{eqnarray}
C^{ds}_{1,\mathcal{G}} & = & \frac{(\lambda^d_{Z,\mathcal{G}})_{sd}^2}{2m_{Z}^2}
+\frac{(\lambda^d_{Z',\mathcal{G}})_{sd}^2}{2m_{Z'}^2}
+\frac{(\lambda^d_{X^0,\mathcal{G}})_{sd}(\lambda^{d\ast}_{X^0,\mathcal{G}})_{ds}}{m_{X^0}^2},\\
C^{ds}_{2,\mathcal{G}} &=& -\sum_\phi\frac{(\Gamma^{d\ast}_{\phi,\mathcal{G}})^2_{ds}}{4m^2_\phi}+\sum_{i=1,2}\frac{(\Gamma^{d\ast}_{A_i,\mathcal{G}})^2_{ds}}{4m^2_{A_i}},\\ 
\widetilde{C}^{ds}_{2,\mathcal{G}} & = & -\sum_\phi\frac{(\Gamma^{d}_{\phi,\mathcal{G}})^2_{sd}}{4m^2_\phi} + \sum_{i=1,2}\frac{(\Gamma^{d}_{A_i,\mathcal{G}})^2_{sd}}{4m^2_{A_i}},\\
C^{ds}_{4,\mathcal{G}} & = & -\sum_\phi\frac{(\Gamma^{d\ast}_{\phi,\mathcal{G}})_{ds}(\Gamma^d_\phi)_{sd}}{2m^2_\phi} - \sum_{i=1,2}\frac{(\Gamma^{d\ast}_{A_i,\mathcal{G}})_{ds}(\Gamma^d_{A_i})_{sd}}{2m^2_{A_i}}.\label{kaon mixing}
\end{eqnarray}
The neutral gauge boson interactions are
\begin{eqnarray}
\mathcal{L}_{\rm gauge} 
&=&\bar u_L ({\lambda}_{Z,\mathcal{G}}^u)\gamma^\mu u_L Z_\mu
+\bar d_L ({\lambda}_{Z,\mathcal{G}}^d)\gamma^\mu d_L Z_\mu
+\bar u_L ({\lambda}_{Z',\mathcal{G}}^u)\gamma^\mu u_L Z'_\mu
\label{gauge boson physical couplings}\\
&+&\bar d_L ({\lambda}_{Z',\mathcal{G}}^d)\gamma^\mu d_L Z'_\mu
+(\bar u_L ({\lambda}_{X^0,\mathcal{G}}^u)\gamma^\mu u_L X^0_\mu+
\bar d_L ({\lambda}_{X^0,\mathcal{G}}^d)\gamma^\mu d_L X^0_\mu+\hc),\nonumber
\end{eqnarray}
where the couplings ${\lambda}^q_{Z,Z',X^0,\mathcal{G}}$ are give in Appendix \ref{gauge boson neutral currents}. The right-handed quarks are in the same representation and therefore they don't have flavour changing couplings with neutral gauge bosons. 
The scalar and pseudo-scalar Yukawa couplings are
\begin{eqnarray}
\mathcal{L}_{\textrm{physical Yukawa}}
&=& \sum_{\phi}\frac{1}{\sqrt{2}}\bar{u}_{L}{\Gamma}^u_{h}u_{R}\phi
+\sum_\phi\frac{1}{\sqrt{2}}\bar{d}_{L}{\Gamma}^d_{h}d_{R}\phi\\
&+&\sum_{i=1,2}\frac{i}{\sqrt{2}}\bar{u}_{L}{\Gamma}^u_{A_\alpha}u_{R}A_i+
\sum_{i=1,2}\frac{i}{\sqrt{2}}\bar{d}_{L}{\Gamma}^d_{A_\alpha}d_{R}A_i+\hc,\nonumber
\end{eqnarray}
where couplings are obtained from Eqs \eqref{up yukawas prime} and \eqref{down yukawas prime} by rotating the primed gauge eigenstate quark fields into mass primeless eigenstates: $\Gamma^u_{h}=U_L^u({\Gamma'}^u_{h})U_{R}^{u\dagger}$ and $\Gamma^d_{h}=U_{L}^d({\Gamma'}^d_{h})U_{R}^{d\dagger}$.
The effective Hamiltonian and Wilson coefficients for other neutral mesons are obtained from Eq. (\ref{kaon mixing}), by  replacing quark flavour indices\footnote{The neutral mesons composed of quarks with different flavour have the following quark content: $K^0=d\bar{s}$, $B_d^0=d\bar{b}$,$B_s^0=s\bar{b}$, $D^0=c\bar{u}$.}.

The mass difference for neutral kaon is 
\be
\Delta m_{K^0} = 2|M_{12}|,\quad \textrm{where} \quad M_{12}=\langle K^0 | \mathcal{H}_{\rm eff}^{K^0\textrm{-}\bar{K}^0}|\bar K^0\rangle.
\ee
Using the hadronic matrix elements found in \cite{Gabbiani:1996hi}, the mass difference becomes 
\be
\Delta m_{K^0} = \frac{2}{3}m_{K} f_K^2\left| C_1^{ds}-(C_2^{ds}+\widetilde{C}_2^{ds})\frac{5}{8}\left(\frac{m_K}{m_d+m_s}\right)^2+C_4^{ds}\left(\frac{1}{8}+\frac{3}{4}\left(\frac{m_K}{m_d+m_s}\right)^2\right)\right|. 
\ee

\begin{table}[]
    \centering
    \begin{tabular}{c|cccc}
        Meson & $K^0$ & $B_d$ & $B_s$ & $D$ \\\hline 
        Mass (MeV) & 497.611 & 5279.66 & 5366.92 & 1864.84\\
        Decay constant (MeV) & 155.6 & 190.9 & 227.2 & 211.9\\
    \end{tabular}
    \caption{Selected neutral meson masses and decay constants \cite{ParticleDataGroup:2022pth}. 
    }
    \label{tab:meson}
\end{table}

The other meson mass differences follow analogously. We use neutral meson masses and decay constants presented in Table~\ref{tab:meson}. 
We study next in detail the coupling structure of all the mediators for each generation assignment. 
We will first look into scalar couplings, paying particular attention to the $125$ GeV Higgs, which on the first glance seems  dangerous scalar mediator due to its relative lightness. After the scalar coupling the gauge boson couplings are investigated.  We present the order of magnitude estimates for all the couplings related to the tree-level neutral meson mixing mediators. These estimates are based on the assumption of CKM-naturality, presented in previous section, and use the quark rotation matrix textures of Eqs. \eqref{up rotation texture} and \eqref{down rotation texture}.  

\subsection{Scalar couplings}
We will first present the exact analytical forms of the scalar coupling, after which we look into order of magnitude estimates based on the quark rotation matrix textures in Eqs. \eqref{up rotation texture} and \eqref{down rotation texture}.
The physical Yukawa couplings of mass eigenstate scalars and pseudo-scalars  are obtained from  Eq. (\ref{up yukawas prime}) and Eq. (\ref{down yukawas prime}) by redefining the quark fields according to Eq. \eqref{quark mass matrix diagonalization}.
For CP-even scalars the physical Yukawa couplings for up type quarks can be written as:
\begin{eqnarray}
(\Gamma^u_{\phi,\mathcal{G}})_{ij} && =  
\sqrt{2}m_j \Bigg\{\frac{U^{H}_{1\phi}}{v_\eta}\delta_{ij}
+\left(\frac{U^{H}_{2\phi}}{v_\rho}-\frac{U^{H}_{1\phi}}{v_\eta}\right) (U_L^u)_{i\mathcal{G}}(U_L^{u\dagger})_{\mathcal{G}j}
+\left(\frac{U^{H}_{3\phi}}{v_\chi}-\frac{U^{H}_{1\phi}}{v_\eta}\right) (U_L^u)_{i4}(U_L^{u\dagger})_{4j}\nonumber\\
&&
+\left(\frac{U^{H}_{4\phi}}{v_\chi}\right)(U_L^u)_{i\mathcal{G}} (U_L^{u\dagger})_{4j}
+\left(\frac{U^{H}_{5\phi}}{v_\rho}\right) (U_L^u)_{i4} (U_L^{u\dagger})_{\mathcal{G}j}\},
\label{scalar up yukawa}
\end{eqnarray}
where $\mathcal{G}$ is the discriminated generation and $\phi=2,3,4,5$ correspond to $h,H_1,H_2,H_3$ respectively. For the pseudo-scalars the Yukawa couplings to up are
\begin{eqnarray}
(\Gamma^u_{A_\alpha,\mathcal{G}})_{ij} & = & -\sqrt{2}m_j  \Bigg\{\frac{U^{A}_{1\alpha}}{v_\eta}\delta_{ij}
+\left(\frac{U^{A}_{2\alpha}}{v_\rho}+\frac{U^{A}_{1\alpha}}{v_\eta}\right) (U_L^u)_{i\mathcal{G}}(U_L^{u\dagger})_{\mathcal{G} j}
+\left(\frac{U^{A}_{3\alpha}}{v_\chi}+\frac{U^{A}_{1\alpha}}{v_\eta}\right) (U_L^u)_{i4}(U_L^{u\dagger})_{4j}\nonumber\\
&&
+\left[\frac{U^{A}_{4\alpha}}{v_\chi}\right] (U_L^u)_{i\mathcal{G}} (U_L^{u\dagger})_{4j}
+\left[\frac{U^{A}_{5\alpha}}{v_\rho}\right] (U_L^u)_{i4} (U_L^{u\dagger})_{\mathcal{G} j}\Bigg\},
\label{A coupling to up}
\end{eqnarray}
where $\alpha=1,2$ denotes the pseudo-scalar in question. 

Here we have used $m^u_{\rm diag} = U^u_L m^u U_R^{d\dagger}$ to eliminate the $U_R^u$ and expressed it only in terms of $U_L^u$. This makes the structure more pleasant, as only the left-handed rotation matrices have  observable effects in the form of CKM matrix. We are mainly interested in flavour violating couplings between SM quarks. From above equations, the term containing only SM quark rotation matrix elements is the second terms on  the first line. This 
seems to be the largest contribution to SM quark off-diagonal couplings as the SM block in the \eqref{up rotation texture} is not suppressed with $\delta$, like the elements involving exotic quarks. The other terms in  \eqref{scalar up yukawa} and \eqref{A coupling to up} contain quark rotation matrix elements involving exotic quarks and are therefore suppressed by at least one factor of $\delta$. For $h$, $H_1$ and $A_1$ the specific forms of scalar and pseudo-scalar rotation matrix elements, \eqref{CP-odd eigenvectors} and \eqref{CP-even rotation}, are crucial in first lines for major cancellations of flavour violating effect. We will go into details regarding this later.  

The down-type quark Yukawa couplings for scalars and pseudo-scalars are analogously to up type quarks
\begin{eqnarray}
&&(\Gamma^d_{\phi,\mathcal{G}})_{ij}  =  \sqrt{2}m_j \Bigg\{\frac{U^{H}_{1\phi}}{v_\eta}\delta_{ij}
+\left(\frac{U^{H}_{2\phi}}{v_\rho}-\frac{U^{H}_{1\phi}}{v_\eta}\right) \left[(U_L^d)_{i \widetilde{\mathcal{G}}_1}(U_L^{d\dagger})_{\widetilde{\mathcal{G}}_1 j}+(U_L^d)_{i\widetilde{\mathcal{G}}_2}(U_L^{d\dagger})_{\widetilde{\mathcal{G}}_2 j}\right]\label{scalar down yukawa}\\
&&+\left[\frac{U^{H}_{3\phi}}{v_\chi}-\frac{U^{H}_{1\phi}}{v_\eta}\right] \left[(U_L^d)_{i4}(U_L^{d\dagger})_{4j}+ (U_L^d)_{i5}(U_L^{d\dagger})_{5j}\right]
\nonumber\\
&&+\left[\frac{U^{H}_{4\phi}}{v_\chi}\right] \left[(U_L^{d})_{i \widetilde{\mathcal{G}}_1}(U_L^{d\dagger})_{4j}
+(U_L^{d})_{i \widetilde{\mathcal{G}}_2}(U_L^{d\dagger})_{5j}\right]
+\left[\frac{U^{H}_{5\phi}}{v_\rho}\right] \left[(U_L^{d})_{i4}(U_L^{d\dagger})_{\widetilde{\mathcal{G}}_1 j}
+(U_L^{d})_{i5}(U_L^{d\dagger})_{\widetilde{\mathcal{G}}_2 j}\right]
\Bigg\},
\nonumber\\
&&
\nonumber
\end{eqnarray} 
and
\begin{eqnarray}
&&(\Gamma^d_{A_\alpha,\mathcal{G}})_{ij}  =  \sqrt{2}m_j\Bigg\{\frac{U^{A}_{1\alpha}}{v_\eta}\delta_{ij}
+\left(-\frac{U^{A}_{2\alpha}}{v_\rho}-\frac{U^{A}_{1\alpha}}{v_\eta}\right)\left[(U_L^d)_{i\widetilde{\mathcal{G}}_1}(U_L^{d\dagger})_{\widetilde{\mathcal{G}}_1 j}+(U_L^d)_{i\widetilde{\mathcal{G}}_2}(U_L^{d\dagger})_{\widetilde{\mathcal{G}}_2 j}\right]\nonumber\\
&&+\left[-\frac{U^{A}_{3\alpha}}{v_\chi}-\frac{U^{A}_{1\alpha}}{v_\eta}\right] \left[(U_L^d)_{i4}(U_L^{d\dagger})_{4j}+ (U_L^d)_{i5}(U_L^{d\dagger})_{5j}\right]
\label{A coupling to down}\\
&&-\left[\frac{U^{A}_{4\alpha}}{v_\chi}\right] \left[(U_L^{d})_{i\widetilde{\mathcal{G}}_1}(U_L^{d\dagger})_{4j}
+(U_L^{d})_{i\widetilde{\mathcal{G}}_2}(U_L^{d\dagger})_{5j}\right]-\left[\frac{U^{A}_{5\alpha}}{v_\rho}\right]\left[(U_L^{d})_{i4}(U_L^{d\dagger})_{\widetilde{\mathcal{G}}_1 j}
+(U_L^{d})_{i5}(U_L^{d\dagger})_{\widetilde{\mathcal{G}}_2 j}\right]\Bigg\}.\nonumber
\end{eqnarray}
Here the $\mathcal{G}$ is the discriminated generation in triplet, and $\widetilde{\mathcal{G}}_1$ and $\widetilde{\mathcal{G}}_2$ are the generations in antitriplet where the former is the lighter generation and the latter the heavier.
The down-type quarks have similar structure of their couplings compared to the up-type quark case in \eqref{scalar up yukawa}. The down-type quarks are slightly more relevant, as they form three neutral mesons, as opposed to up-type quarks, who form only one neutral meson composed of quarks of different flavour.

The $125$ GeV Higgs is potentially dangerous FCNC mediator due to its relative lightness: it is at least order of magnitude lighter than the BSM FCNC mediators. We will next estimate the order of magnitude of scalar and pseudo-scalar mediator flavour violating couplings. 
We obtain the order of magnitude estimates for its couplings to up- and down-type quarks from \eqref{scalar up yukawa} and \eqref{scalar down yukawa}  by using $125$ GeV Higgs eigenvector components from Eq. \eqref{cp-even rotation matrix estimate} with quark rotation matrix textures \eqref{up rotation texture} and \eqref{down rotation texture}.
The up-type coupling matrix is
\begin{eqnarray} 
&\Gamma_{h,\mathcal{G}}^u &\sim \left(\begin{array}{cccc}
y_u &
y_c\delta^2\epsilon^{a_{\mathcal{G}}} &
y_t\delta^2\epsilon^{b_{\mathcal{G}}} &
\frac{m_U}{v_\chi}\epsilon^{3-L_u}\\
\times &
y_c &
y_t\delta^2\epsilon^{2(1-L_u)} &
\frac{m_U}{v_\chi}\epsilon^{2-L_u}\\
\times &
\times &
y_t &
\frac{m_U}{v_\chi}\epsilon^{-L_u}\\
\times &
\times &
\times &
\frac{m_U}{v_\chi}\delta\epsilon^{-2L_u}
\end{array}\right),
\end{eqnarray}
where $a_{1st}=a_{2nd}=1$, $a_{3rd}=5-2L_u$, $b_{1st}=b_{2nd}=3$, $b_{3rd}=3-2L_u$. The order of magnitude estimate is the same in the cases of discriminates 1st and 2nd generation.    
The down-type coupling matrix is
\begin{eqnarray}
&\Gamma_{h,\mathcal{G}}^d &\sim\left(\begin{array}{ccccc}
y_d &
y_s\delta^2\epsilon^{} &
y_b\delta^2\epsilon^{3} &
\frac{m_{D_1}}{v_\chi}\epsilon^{3-L_{D}} &
\frac{m_{D_2}}{v_\chi}\epsilon^{3-L_{D}}\\
%%%%%%%%
\times &
y_s &
y_b\delta^2\epsilon^{2} &
\frac{m_{D_1}}{v_\chi}\epsilon^{2-L_{D}} &
\frac{m_{D_2}}{v_\chi}\epsilon^{2-L_{D}}\\
%%%%%%%%
\times &
\times &
y_b &
\frac{m_{D_1}}{v_\chi}\epsilon^{-L_{D}} &
\frac{m_{D_2}}{v_\chi}\epsilon^{-L_{D}}\\
%%%%%%%%
\times &
\times & 
\times &
\frac{m_{D_1}}{v_\chi}\delta &
\frac{m_{D_2}}{v_\chi}\epsilon^{-2L_D}\\
%%%%%%%% 
\times & 
\times &
\times &
\frac{m_{D_1}}{v_\chi}\delta\epsilon^{-2L_{D}} &
\frac{m_{D_2}}{v_\chi}\delta
\end{array}\right),
\end{eqnarray}
which is the same for all generation assignments. 
We have omitted the terms below the diagonal as they are proportional to smaller Yukawa couplings than above the diagonal as can be easily seen from Eqs (\ref{scalar up yukawa}) and (\ref{scalar down yukawa}).

For the down-sector all relevant flavour violating couplings for all generation assignments are the same, at least in their order of magnitude. This mediator cannot be used to distinguish between generation assignments in meson mixing. For the up-sector the only relevant coupling  in meson mixing is between up and charm, where the discriminated third generation differs from other assignments. 

For all generation assignments the diagonal couplings of SM quarks are those of the SM, up to miniscule corrections and the flavour violating couplings amongst the SM quarks are heavily suppressed by $\delta^2$, yielding to their decoupling if $SU(3)_L$-breaking scale is large. The diagonal couplings of exotic quarks $U$, $D_1$ and $D_2$ also decouple with large $v_\chi$. This is desirable, as this way the exotic quarks  do not contribute to the $125$ GeV Higgs production in hadron colliders through gluon-gluon fusion. The flavour violating couplings of $U$ to SM quarks, however, are potentially large and do not experience decoupling. This could potentially yield to interesting signals at the LHC or future hadron colliders.

The $125$ GeV Higgs and other heavier scalars and pseudo-scalars do not give strong contribution to neutral meson mixing.  This is due to their flavour violating couplings among SM quarks being subject to suppression from multiple sources:
\begin{itemize}
\item \textbf{125 GeV Higgs}: The 125 GeV Higgs couplings are suppressed due to almost total cancellation of $U^H_{12}$ and $U^H_{22}$, as seen in equations \eqref{scalar up yukawa} and \eqref{scalar down yukawa} with Eqs. \eqref{UH12} and \eqref{UH22}, leading into $\delta^2$-suppression\footnote{In \cite{Huitu:2019kbm} the flavour violating couplings between SM quarks and Higgs were found to be suppressed only by $\delta$. This is due to the presence of two $SU(3)_L$-breaking VEVs.}. This suppression takes place without any tuning of parameters. 
On top of that, additional suppression is provided small SM Yukawa couplings and powers of $\epsilon$, that originate from pure SM element in left-handed quark rotation matrices. In total the contribution from $125$ GeV Higgs to the Wilson coefficients is: $C^h_2\sim y_q^2 \delta^2\epsilon^p/v_\chi^2$, where $p$ is small positive integer.
\item \textbf{H}$_\mathbf{1}$\textbf{-A}$_\mathbf{1}$ \textbf{-- cancellation}: The contributions from heavy scalar $H_1$ and heavy pseudo-scalar $A_1$ almost completely cancel: $H_1$ and $A_1$ are nearly mass degenerate, as seen from Eqs. \eqref{A1 and A2 mass} and \eqref{H1 mass}. Their couplings are also almost identical in magnitude, which can be seen by using  \eqref{CP-odd eigenvectors} and \eqref{cp-even rotation matrix estimate} in Eqs. \eqref{scalar up yukawa}, \eqref{A coupling to up}, \eqref{scalar down yukawa} and \eqref{A coupling to down}.
There is a sign difference between their contributions due to pseudo-scalar nature of $A_1$. This leads into almost complete cancellation between their contributions. The combined contribution from them is $C^{H_1}_2+C^{A_1}_2\sim y_q^2 \delta^2\epsilon^p/v_\chi^2$.
%%%%%%%%%%%%%%%%%%%%%%%%%%%%%%%%%%%%%%%%%%%%%
\item \textbf{Small} $\mathbf{U^H}$ \textbf{and} $\mathbf{U^A}$ \textbf{elements}: The flavour violating couplings of $H_2$, $H_3$ and $A_2$ to SM quarks are  heavily suppressed. This can be seen from scalar \eqref{cp-even rotation matrix estimate}  and pseudo-scalar eigenvectors \eqref{CP-odd eigenvectors}: The main contribution to pure SM flavour violating couplings originates from second line in Eqs. \eqref{scalar up yukawa}, \eqref{A coupling to up}, \eqref{scalar up yukawa}, \eqref{A coupling to down}. The relevant scalar  eigenvector components to $H_2$ and $H_3$ are suppressed by $\delta$ and $\delta^2$ respectively, yielding into heavy suppression when their heavy mass is taken into account. The contribution from pseudo-scalar $A_2$ is completely negligible as the would-be main contributions to pure SM flavour violation vanish, due to many of its eigenvector components being zero.
\end{itemize}
All in all the scalar and pseudo-scalar contributions to neutral meson mixing is heavily suppressed due to structure of the scalar mass matrices. These effects are always present without any tuning of parameters. These suppression effects are independent of the chosen texture in the quark mass matrices, and are therefore present if different textures are used in the quark sector.

\subsection{Gauge boson coupling textures}

\subsubsection{$Z$ boson textures}
We can obtain estimates for the flavour violating quark couplings of $Z$  by using quark rotation matrix textures \eqref{up rotation texture} and \eqref{down rotation texture} (see Appendix \ref{gauge boson neutral currents} for details). Below we present the only the off-diagonal couplings for each generation assignment,
\be\label{Z up texture 1st}
\lambda_{Z,\mathcal{G}}^u\sim\left(\begin{array}{llll}
*~~~~~ & 
\delta^2\epsilon^{5-2L_u}+\epsilon^{a_{\mathcal{G}}}\theta &
\delta^2\epsilon^{3-2L_u}+\epsilon^3\theta & 
\delta\epsilon^{3-L_u}\\
\times &
 * &
 \delta^2\epsilon^{2-2L_u}+\epsilon^{b_{\mathcal{G}}}\theta &
 \delta\epsilon^{2-L_u} 
\\
 \times &
 \times &
 * & 
 \delta\epsilon^{2-L_u}
\\
\times &
\times &
\times &
 *
\end{array}\right),
\ee
where $a_{1st}=a_{2nd}=1$, $a_{3rd}=5$, $b_{1st}=4$, $b_{2nd}=b_{3rd}=2$.
The coupling matrices are symmetric and the elements below the diagonal are presented by crosses. We are only interested in flavour violating effects, and the diagonal couplings that are represented by an asterisk. We employ the same notation for the other gauge boson coupling textures as well.   For the meson mixing only the coupling between up and charm quarks is relevant. It is essentially the same for all generation assignments and cannot therefore be used to distinguish the discriminated generation for Z-mediation.

In the case of down-type quarks  we assume for simplicity that  $L_{D_1}=L_{D_2}\equiv L_D$. The $L_{D_1}$ and $L_{D_2}$ mainly affect the mixing between SM quarks and exotic quarks, and are therefore not relevant when considering the BSM contribution to neutral meson mass difference\footnote{The 1-loop contributions to neutral meson mixing coming from exotic quarks was studied in detail in \cite{Huitu:2019kbm}.}.  In this case the coupling textures are,
\begin{eqnarray}
\lambda_{Z,\mathcal{G}}^d &\sim&\left(\begin{array}{lllll}
*~~~~~ & 
\epsilon(\delta^2\epsilon^{4-2L_{d}}+\epsilon^{a_{\mathcal{G}}}\theta) &
\epsilon^3(\delta^2\epsilon^{-2L_{d}}+\theta) & 
\delta\epsilon^{3-L_{d}} &
\delta\epsilon^{3-L_{d}}\\
\times &
* &
\epsilon^2(\delta^2\epsilon^{-2L_{d}}+\epsilon^{b_{\mathcal{G}}}\theta)&
\delta\epsilon^{2-L_{d}}&
\delta\epsilon^{2-L_{d}} \\
\times &
\times &
 * & 
\delta\epsilon^{-L_{d}} &
\delta\epsilon^{-L_{d}}\\
 \times &
 \times &
 \times &
 * &
 \delta^2\epsilon^{-2L_{d}}\\
 \times &
 \times &
 \times &
 \times &
 *
\end{array}\right),\label{Z down texture 1st}
\end{eqnarray}
where $a_{1st}=a_{2nd}=0$, $a_{3rd}=4$, $b_{1st}=2$, $b_{2nd}=b_{3rd}=0$. 
For the mesons composed of down-type quarks all the generation assignments  predict essentially the same contribution for neutral meson mass difference from $Z$ boson mediation. Also similarly to up-type quarks,  the mixing between SM quarks is essentially the same for all generation assignments. 

The contribution to Wilson coefficients, relevant to neutral meson mass difference, from $Z$ boson is heavily suppressed. 
The $Z$ boson has two sources of flavour violation that are visible in coupling matrices before flavour rotation in  Eqs. \eqref{Z and Zprime up before} and \eqref{Z and Zprime down before}: the SM quark couplings in the diagonal are mostly the same, except for the small extra term in the element corresponding to the discriminated generation. This term originates from $Z\textrm{-} Z'$ mixing and is proportional to $\theta$. Second contribution is from exotic quark mixing with the SM quarks: the exotic quark couplings in  Eqs.~\eqref{Z and Zprime up before} and \eqref{Z and Zprime down before} are different compared to SM couplings in the diagonal. When performing the flavour rotation into physical couplings, the quark rotation matrix elements away from the SM $3\times 3$-block pick out this difference. As a result the contribution from the mixing of SM quark and exotic quarks is proportional to $\delta^2$. These contributions are clearly visible in Eq.~\eqref{Z up texture 1st}, for the $Z$-coupling order of magnitude estimates. 
The contribution from the $Z$ to the Wilson coefficients is $C^Z_1\sim g_3^2\delta^2\epsilon^p/v_\chi^2$ ($p$ is non-negative integer), which is slightly larger than that of $125$~GeV Higgs due to gauge couplings being larger than relevant Yukawa couplings (top does not hadronize).

\subsubsection{$Z'$ textures}
The order of magnitude estimates for $Z'$ couplings to up- and down-type quarks are
\be\label{Zprime texture}
\lambda_{Z',\mathcal{G}}^u\sim\left(\begin{array}{llll}
*~~ & 
\epsilon^{a_{\mathcal{G}}} &
\epsilon^3 & 
\delta\epsilon^{3-L_u}\\
\times &
 * &
 \epsilon^{b_{\mathcal{G}}} &
 \delta\epsilon^{2-L_u}\\
 \times &
 \times &
 * & 
 \delta^{-L_u}
\\
 \times &
 \times &
 \times &
 *
\end{array}\right),\quad
\lambda_{Z',\mathcal{G}}^d\sim\left(\begin{array}{lllll}
*~~ & 
\epsilon^{a_{\mathcal{G}}}~~ &
\epsilon^3~~ & 
\delta\epsilon^{3-L_{d}} &
\delta\epsilon^{3-L_{d}}\\
\times &
*~~ &
\epsilon^{b_{\mathcal{G}}} &
\delta\epsilon^{2-L_{d}}&
\delta\epsilon^{2-L_{d}} \\
 \times &
 \times &
 * & 
\delta\epsilon^{-L_{d}} &
\delta\epsilon^{-L_{d}}\\
 \times &
 \times &
 \times &
 * &
 \delta^2\epsilon^{-2L_{d}}\\
 \times &
 \times &
 \times &
 \times &
 *
\end{array}\right),
\ee
where $a_{1st}=a_{2nd}=1$, $a_{3rd}=5$, $b_{1st}=4$, $b_{2nd}=b_{3rd}=2$.
There are clear differences between different generation assignments, as opposed to Z-mediation.
For both SM up and down quarks the flavour violating couplings are not suppressed by $\delta$, like for all the scalars and the $Z$. The origin of this is in that the discriminated generation has different coupling compared to other generations before flavour rotation, as elaborated in the Appendix \ref{gauge boson currents}. The contribution to the Wilson coefficient from $Z'$ is: $C^{Z'}_1\sim g_3^2\epsilon^q/v_\chi^2$. The scalars and $Z$ have a factor of $\delta^2$ suppression compared to this, making $Z'$ dominant, except at low $SU(3)_L$-breaking scale  $\lesssim 10$ TeV, where $Z$ can dominate for certain generation assignments and mesons. 

For neutral kaon mixing (the $12$-element in down-type coupling), we find that the discriminated third generation provides significant suppression by the factor $\sim \epsilon^4$ compared to the other generation assignments. This seems very appealing choice for model building when one's aim is to avoid neutral meson mixing constraints, as the kaon bounds are the most stringent. For $B_d$ meson mixing all the generation assignments produce the same contribution. 
Similar to down sector, the discriminated third generation professes suppression in $12$-element for up-type coupling compared to other generation assignments. The $13$- and $23$-elements are the same for all generation assignments in the case of up, which is, however, not that relevant as the top quark does not hadronize.  
All generation assignments predict essentially the same couplings between SM quarks and exotic quarks. One also notes that the couplings of SM quarks to exotic quarks are suppressed by factor $\delta$ and are therefore small. It is therefore unlikely that these couplings could be utilized to distinguish 331-$Z'$ from a generic $Z'$ in future colliders.  

\subsubsection{$X^0$ textures}
The order of magnitude of the up-type quark couplings to $X^0$ are,
\be\label{X0 up texture}
\lambda_{X^0,\mathcal{G}}^u\sim\left(\begin{array}{llll}
* & 
\delta\epsilon^{a^u_\mathcal{G}-L_u} &
\delta\epsilon^{b^u_\mathcal{G}-L_u} & 
\epsilon^{b^u_\mathcal{G}}\\
%%%%%%%%%%%%%%%%%%%%%%%
\delta\epsilon^{\widetilde{a}^u_\mathcal{G}-L_u} & 
* &
\delta\epsilon^{c^u_\mathcal{G}-L_u} & 
\epsilon^{c^u_\mathcal{G}}\\
%%%%%%%%%%%%%%%%%%%%%%%
\delta\epsilon^{\widetilde{b}^u_\mathcal{G}-L_u} & 
\delta\epsilon^{\widetilde{c}^u_\mathcal{G}-L_u} &
* & 
\epsilon^{d^u_\mathcal{G}}\\
%%%%%%%%%%%%%%%%%%%%%%%
\delta^2\epsilon^{\widetilde{b}^u_\mathcal{G}-2L_u} & 
\delta^2\epsilon^{\widetilde{c}^u_\mathcal{G}-2L_u} &
\delta^2\epsilon^{d^u _\mathcal{G}-2L_u} & 
*\\
\end{array}\right),
\ee
and for down-type quarks,
\be\label{X0 down texture}
\lambda_{X^0,\mathcal{G}}^d\sim\left(\begin{array}{ccccc} 
* & 
\delta\epsilon^{\widetilde{a}^d_\mathcal{G}-L_d} &
\delta\epsilon^{\widetilde{b}^d_\mathcal{G}-L_d} & 
\delta^2\epsilon^{\widetilde{b}^d_\mathcal{G}-2L_d} &
\delta^2\epsilon^{\widetilde{b}^d_\mathcal{G}-2L_d}\\
%%%%%%%%%%%%%%%%%%%%%%%
\delta\epsilon^{a^d_\mathcal{G}-L_d}  & 
* &
\delta\epsilon^{\widetilde{c}^d_\mathcal{G}-L_d} & 
\delta^2\epsilon^{\widetilde{c}^d_\mathcal{G}-2L_d} &
\delta^2\epsilon^{\widetilde{c}^d_\mathcal{G}-2L_d}\\
%%%%%%%%%%%%%%%%%%%%%%%
\delta\epsilon^{b^d_\mathcal{G}-L_d} & 
\delta\epsilon^{c^d_\mathcal{G}-L_d}  &
* & 
\delta^2\epsilon^{d^d_\mathcal{G}-2L_d} &
\delta^2\epsilon^{d^d_\mathcal{G}-2L_d}\\
%%%%%%%%%%%%%%%%%%%%%%%
\epsilon^{b^d_\mathcal{G}} &
\epsilon^{c^d_\mathcal{G}} &
\epsilon^{d^d_\mathcal{G}} &
* &
\delta\epsilon^{d^d_\mathcal{G}-L_d}\\
%%%%%%%%%%%%%%%%%%%%%%%
\epsilon^{b^d_\mathcal{G}} &
\epsilon^{c^d_\mathcal{G}} &
\epsilon^{d^d_\mathcal{G}} &
\delta\epsilon^{d^d_\mathcal{G}-L_d} &
*
\end{array}\right).
\ee
The powers of $\epsilon$ are given in Table \ref{tab:X powers}. We have included all the couplings as the neutral meson mixing involves couplings from both sides of the diagonal.  
\begin{table}[h]
    \centering
    \begin{tabular}{c|ccccccc|ccccccc}
$\mathcal{G}$  & $a^u_\mathcal{G}$ & $b^u_\mathcal{G}$ & $c^u_\mathcal{G}$ & $d^u_\mathcal{G}$ & $\widetilde{a}^u_\mathcal{G}$ & $\widetilde{b}^u_\mathcal{G}$ & $\widetilde{c}^u_\mathcal{G}$ 
      & $a^d_\mathcal{G}$ & $b^d_\mathcal{G}$ & $c^d_\mathcal{G}$ & $d^d$ & $\widetilde{a}^d_\mathcal{G}$ & $\widetilde{b}^d_\mathcal{G}$ & $\widetilde{c}^d_\mathcal{G}$\\\hline 
$1$st & 2     & 0     & 1     & 3     & 4                 & 6                 & 5
      & 3     & 1     & 0     & 0     & 3                 & 3                 & 2\\
$2$nd & 3     & 1     & 0     & 2     & 3                 & 5                 & 4
      & 2     & 0     & 1     & 0     & 4                 & 3                 & 2\\
$3$rd & 5     & 3     & 2     & 0     & 5                 & 3                 & 2
      & 2     & 0     & 0     & 2     & 3                 & 5                 & 4
    \end{tabular}
    \caption{The powers for different generation assignments in Eqs. (\ref{X0 up texture}) and (\ref{X0 down texture})}
    \label{tab:X powers}
\end{table}

The flavour violating couplings between SM quarks are suppressed by a factor of $\delta$. This originates from $X^0$-couplings before the flavour rotation, that can be seen in Eqs. (\ref{x u before rotation}) and (\ref{x down before rotation}), where $X^0$  SM quarks couple only exotic quarks and not to themselves. The couplings between SM quarks are generated form the exotic edges rotation matrices (\ref{up rotation texture}) and (\ref{down rotation texture}).  
The contribution to the Wilson coefficient from $X^0$ is: $C^{X^0}_1\sim g_3^2(\epsilon^q\delta)^2/v_\chi^2$. 
Even though it seems at the first glance that the $X^0$ contribution is always smaller than that of $Z'$, the $X^0$ can dominate neutral meson mass difference for certain mesons and generation assignments at low values of $v_\chi$. This is due to involvement of different rotation matrix elements in different cases. We will elaborate on this in  the next section.

\section{Predictions for different generation assignments}\label{numerics}
In the previous section we derived the order of magnitude estimates for the quark flavour violating couplings. 
In this section we compute the BSM contribution to neutral meson mass difference and estimate the  how low the $SU(3)_L$-breaking scale can be for different generation assignments. We also look into LHC  bounds in the case of low BSM scale. 

\subsection{Estimates for neutral meson mixing}

In the previous section we assumed that both up- and down-sector left-handed rotation matrix elements have the same order of magnitude as CKM matrix, in order to get handle on the magnitude of the physical couplings.
However, even though the quark mass matrices respect the CKM-naturality, outlined in Section \ref{CKM naturality}, the rotation matrix entries can deviate significantly from the rough estimates of Eqs. (\ref{up rotation texture}) and (\ref{down rotation texture}). Small rotation matrix elements  would allow for a smaller $SU(3)_L$-breaking scale, providing better opportunities for collider searches.  
This is attractive, but there is a caveat: the CKM matrix links the up- and down-sectors together and therefore both up- and down-sector rotation matrices cannot have small entries in the off-diagonal simultaneously, as in this case CKM matrix would not be realized. By small off-diagonal entries we here mean smaller than estimates of Eqs. (\ref{up rotation texture}) and (\ref{down rotation texture}).   

\begin{figure}
    
    \includegraphics[width=0.95\textwidth]{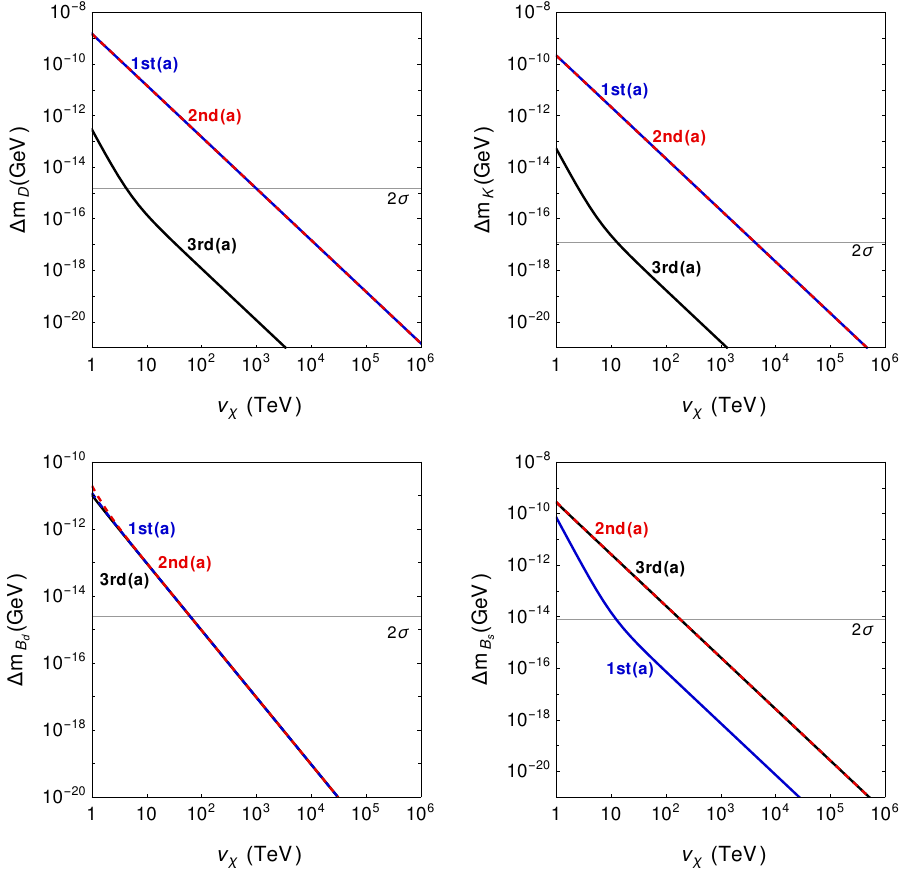}
    \caption{
    The estimates of contribution to neutral meson mass difference as a function of $SU(3)_L$-breaking VEV, using quark rotation matrices of Eqs. (\ref{up rotation texture}) and (\ref{down rotation texture}) with parameters $v_\eta=v_\rho=v/\sqrt{2}$, $L_u=1$, $L_{D_1}=L_{D_2}=-1$. The blue curve corresponds to discriminated first generation, red to second and black to third. The vertical line is the $2\sigma$ bound on corresponding neutral meson mass difference, area above that is excluded.
    }
    \label{meson mass difference from texture}
\end{figure}

\begin{figure}
    
    \includegraphics[width=0.95\textwidth]{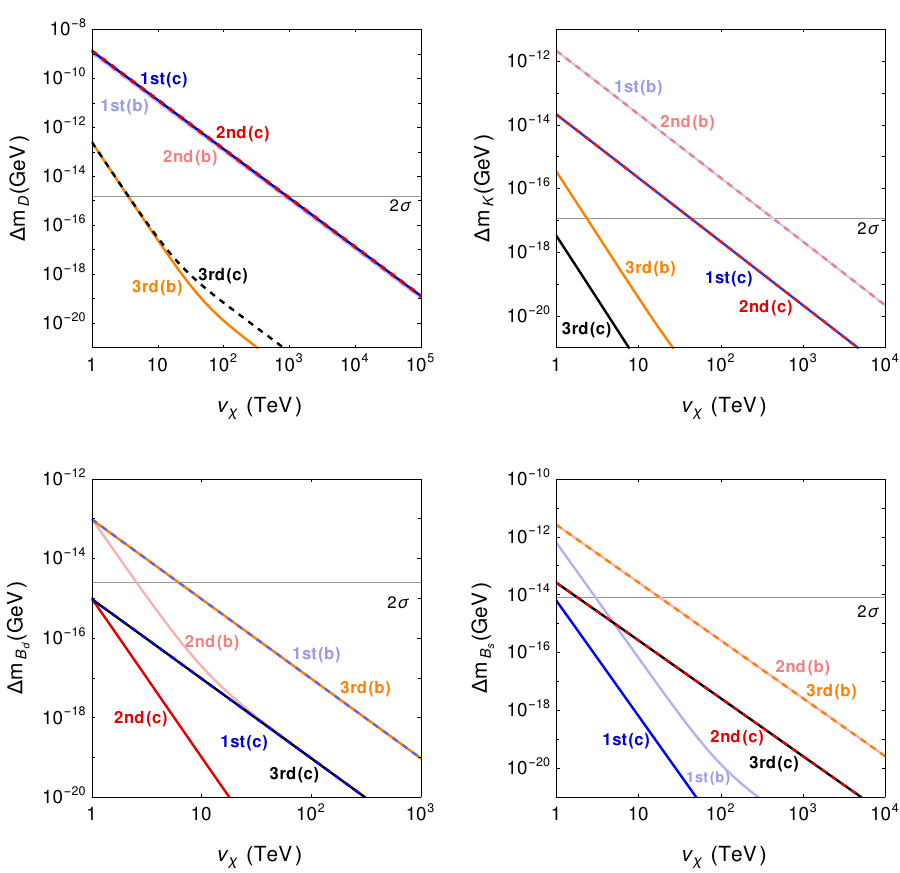}
    \caption{
    The estimates of contribution to neutral meson mass difference as a function of $SU(3)_L$-breaking VEV, using quark rotation matrices of Eq. (\ref{rotation with coefficients}) with parameters $v_\eta=v_\rho=v/\sqrt{2}$, $f=v_\chi$, $L_u=1$, $L_{D_1}=L_{D_2}=-1$ for all three possible generation assignments. Cases \textbf{b} and \textbf{c} correspond to $\beta=0.1$ and $\beta=0.01$ respectively The vertical line is the $2\sigma$ bound on corresponding neutral meson mass difference, area above that is excluded.
    }
    \label{meson mass difference from coefficients in texture}
\end{figure}

In order to explore the effect of smaller rotation matrix elements we parametrize the rotation matrices as:
\be\label{rotation with coefficients}
|U_L^u| = \left(\begin{array}{cccc}
1 & \alpha_1\epsilon & \alpha_2\epsilon^3 & \delta\epsilon^{4}\\
\widetilde{\alpha}_1\epsilon & 1 & \alpha_3\epsilon^2 & \delta\epsilon^{3}\\
\widetilde{\alpha}_2\epsilon^3 & \widetilde{\alpha}_3\epsilon^2 & 1 & \delta\epsilon^{}\\
\delta\epsilon^{4} & \delta\epsilon^{3} & \delta\epsilon^{} & 1
\end{array}\right),\quad
|U_L^d| = \left(\begin{array}{ccccc}
1 & \beta\epsilon & \beta\epsilon^3 & \beta\delta\epsilon^{4} & \beta\delta\epsilon^{4}\\
\beta\epsilon & 1 & \beta\epsilon^2 & \beta\delta\epsilon^{3} & \beta\delta\epsilon^{3}\\
\beta\epsilon^3 & \beta\epsilon^2 & 1 & \beta\delta\epsilon^{} & \beta\delta\epsilon^{}\\
\beta\delta\epsilon^{4} & \beta\delta\epsilon^{4} & \beta\delta\epsilon^{} & 1 & 1\\
\beta\delta\epsilon^{4} & \beta\delta\epsilon^{3} & \beta\delta\epsilon^{} & \delta  & 1
\end{array}\right),
\ee
where $\alpha$, $\widetilde{\alpha}$ and $\beta$ are free parameters. We fix  the parameter $\beta$ and solve for $\alpha$ and $\widetilde{\alpha}$ so the experimental values for CKM matrix are realized. We present three different cases. In the case \textbf{a} we use the rotation matrices of Eqs. (\ref{up rotation texture}) and (\ref{down rotation texture}). This gives us a rough estimate for contribution to the meson mixing for large flavour violating couplings. In the cases \textbf{b} and \textbf{c} we use rotation matrices of Eq. (\ref{rotation with coefficients}) with $\beta=0.1$ and $\beta=0.01$ respectively. These provide smaller entries in the off diagonal for the down sector, while $U_L^u$ remains CKM-like. 
The contribution to neutral meson mass differences in case \textbf{a} is presented in Figure \ref{meson mass difference from texture} and in Figure \ref{meson mass difference from coefficients in texture} for cases \textbf{b} and \textbf{c}. We take into account only the neutral gauge boson mediators and ignore the scalars and pseudo-scalars as their contribution is orders of magnitude smaller than from  gauge bosons. 

The case \textbf{a} represents larger flavour violating couplings as both $U_L^u$ and $U_L^d$ are close to CKM. One can see in Figure \ref{meson mass difference from texture} the difference between the generation assignments, which can also be seen in Eqs. (\ref{Zprime texture}), (\ref{X0 up texture}) and (\ref{X0 down texture}). Taking into account all neutral mesons, the discriminated third generation can be realized with the lowest $SU(3)_L$-breaking scale. 
In the case \textbf{a} the $SU(3)_L$-breaking scale is pushed over $100$ TeV in the case of discriminated third generation and close to $10^4$ TeV for discriminated first and second generation. Even the more hopeful case of discriminated third generation would be beyond the reach of currently planned future colliders. 

The cases \textbf{b} and \textbf{c} show how the required $SU(3)_L$-breaking scale can be lowered if the down sector rotation matrix has lower off-diagonal entries than the CKM. As seen in the Figure \ref{meson mass difference from coefficients in texture} the $SU(3)_L$-breaking scale can be brought as low as $\sim 10$ TeV if the off-diagonal entries in the down sector rotation matrix are suppressed by the factor of $\sim 0.1$ compared to corresponding entries in the CKM. When the down sector rotation matrix is brought closer to unit matrix, the contribution to $\Delta m_K$, $\Delta m_{B_d}$ and $\Delta m_{B_s}$ is reduced, but the contribution to $\Delta m_D$ is not. The discriminated third generation  has suppression in the couplings relevant to D-meson, so the BSM contribution is low and it doesn't impose strong bound. It seems that in order to have the $SU(3)_L$-breaking scale within the collider reach, the third  generation has to be the discriminated one.

For all cases \textbf{a}, \textbf{b} and \textbf{c}, $Z'$ dominates as a mediator for $SU(3)_L$-breaking scales $\gtrsim 10$ TeV, but for lower BSM scale $Z$ and $X^0$ can dominate in certain cases. These cases can be seen as a steeper slope in Figure \ref{meson mass difference from texture}.  In the case of D-meson it is the $Z$ that dominates at low $SU(3)_L$-breaking scale for the third generation. This is due to large suppression in terms of powers of $\epsilon$ for $Z'$ when compared to $Z$, even though $Z$ is suppressed by a factor of $\delta^2$ compared to $Z'$. 
In a similar manner $X^0$ briefly  dominates in the case of discriminated third generation for kaon, for second generation in $B_d$ meson and for first generation in $B_s$ meson.

\subsection{Collider constraints and benchmark points} 
\label{benchmark}

The neutral meson mixing allows the $SU(3)_L$-breaking scale to be as low as $\sim 10$ TeV for discriminated third generation. It will then be probably observable at the LHC. We present here three numerical benchmark scenarios  with $SU(3)_L$-breaking VEV $v_\chi=15$ TeV and electroweak scale VEVs $v_\eta=v_\rho=v/\sqrt{2}$. The parameters of the Yukawa sector 
are given in Appendix \ref{numerical example}. Here we check that the benchmark scenarios are safe from the current constraints on the production exotic up-type quark ($U$), and the two exotic down-type quarks ($D_1$ and $D_2$) 
in addition to the exotic gauge bosons $V^{\pm}$, $Z^{\prime}$ and the non-hermitian boson $X^0$.\footnote{We concentrate here on the exotic quarks and gauge bosons as they are relevant for the neutral meson mixing and further the scalar sector. The $125$ GeV Higgs is very much SM-like as seen in Section \ref{scalar masses}.}  
For our benchmarks the exotic gauge boson masses are $m_{Z'}=5.751$ TeV %5751$ GeV
and $m_{X}=m_V=4.724$ TeV, %4724$ GeV
and the masses of the exotic quarks are given in Table \ref{tab:BPs}. All the benchmarks pass the bounds from neutral meson mixing only for discriminated third generation. The down-type quark rotation matrices $U_L^d$ for these benchmarks resemble those of cases \textbf{b} and \textbf{c} of previous subsection, allowing them to have low BSM contribution to neutral meson mass difference\footnote{The rotation matrices for benchmarks are presented in \ref{BP1}, \ref{BP2} and \ref{BP3}.}.

\begin{table}[h]
    \centering
    \begin{tabular}{c|ccc}
         &\textbf{BP1} & \textbf{BP2} & \textbf{BP3}  \\\hline 
        $m_U$/TeV           & 1.362 & 2.368 & 1.829\\
        $m_{D1}$/TeV        & 2.150 & 6.931 & 9.076\\
        $m_{D2}$/TeV        & 16.007 & 36.369 & 26.922
    \end{tabular}
    \caption{The masses of exotic gauge bosons for the benchmark points. The Yukawa sector parameters are given in Appendix \ref{numerical example}.}
    \label{tab:BPs}
\end{table}

Given the coupling structure, the exotic up-type quark ($U$) has three dominant two-body decay modes. The decay branching ratios are comparable in top quark ($t$) - SM Higgs ($h$) and bottom quark ($b$) - W boson ($W$) states. These are the two most dominant decay modes of $U$ followed by a third channel comprising of top quark and Z boson ($Z$). Direct search of such exotic quarks in the context of the LHC \cite{CMS:2022fck, ATLAS:2024gyc} puts a strong constraint on the chosen mass of $U$. The CMS collaboration reports a stringent limit on the $U$ mass, $m_{U} > 1.5$ TeV from 13 TeV data at a luminosity of 138 fb$^{-1}$, assuming the particle being produced through gluon-gluon initial state decays 50\% of the time into $bW$ exclusively with the other 50\% distributed uniformly between $th$ and $tZ$ channels \cite{CMS:2022fck}.  The results do not change significantly if the $U$ state decays uniformly into $th$ and $tZ$ states only. On the other hand if $U$ decays entirely into a $bW$ final state, the limit slightly improves, $m_{U} > 1.6$~TeV \cite{ATLAS:2024gyc}. Although our benchmark points do not produce exactly same branching ratios as assumed in these simplified scenarios, we keep $m_{U} > 1.5$ TeV as a conservative choice. We have further cross-checked with the obtained branching ratios in our benchmark points that the resultant rate of the leptonic final states  studied at the LHC remains below the observed rate and thus safe from the constraint.

Similarly, for the down-type exotic quarks the experimental study concentrate on three decay modes, namely, $tW$, $bh$ and $bZ$ \cite{CMS:2022fck,ATLAS:2024gyc}. Limit on the masses of such particles are strongest when the dominant decay mode of the down-type quark is $tW$ with a 50~\% branching ratio with the other 50~\% being shared uniformly by the two other decay modes, $bh$ and $bZ$. In such cases, the mass limit extends to nearly 1.5 TeV. In contrast, if the particle is assumed to decay uniformly into only $bh$ and $bZ$ states, the mass limit on the down-type quarks come down to $\sim$ 1.1 TeV \cite{CMS:2022fck}. In our present scenario, however, we require $D_1$ and $D_2$ to be heavier than some new bosons ($V^{\pm}$, $X^{0}$) and therefore, can decay via these gauge bosons as well associated with light quarks. This in fact happens to be the dominant scenario. The new bosonic states further decays into SM particles resulting in higher jet multiplicity in the final states arising from pair production of the exotic quarks. A detailed collider analysis is needed to ascertain the LHC sensitivity to such scenarios. However, in all our benchmark points, $D_2$ is outside the reach of LHC energy whereas masses of $D_1$ in two of the benchmark points are outside LHC reach and one of them has $m_{D_1}\sim 2150$ GeV which is safely above the existing limit.  

In the present scenario, the heavy $V^\pm$ almost entirely ($\sim$ 90-95\%) decays through the exotic quarks in association with light quarks while the rest of the time it decays into SM quarks. In such cases, the limit on $V^\pm$ masses are weaker, $\sim$ 3-3.3 TeV \cite{CMS:2022tdo}. The $Z^{\prime}$ decays through exotic quarks $\sim$ 30-40\% of the time and the  rest of the time it decays through SM quark pairs. For the cases when a $Z^{\prime}$ decays through vector-like quarks, the limit on the $Z^{\prime}$ production cross-section is in the pb order \cite{CMS:2017hwg,CMS:2018wkq}. For our benchmark points this cross-section is $\sim$ fb, which makes them unobservable with the existing LHC data. As a result of the combined effect of low branching ratios into SM quarks and small production cross-section, the $Z^{\prime}$ masses in our benchmark points can be explored only at high luminosity LHC \cite{CMS:2024lwn,ATLAS:2018tvr}. The $X^0$-boson again mostly decays through exotic quarks and the masses chosen for our benchmark points can only be explored at high luminosity LHC.

\section{Conclusions}
\label{conclusions}

We found that the choice of discriminated quark generation greatly affects the flavour violating couplings in a generic 331-model with $\beta=-1/\sqrt{3}$.
We took into account all the tree-level mediators of neutral meson mixing.
The flavour violating couplings depend crucially on the diagonalization matrices of quarks, whose structure is determined by the mass matrices.  We made the natural ansatz for the structure of the quark mass matrices which produces the CKM matrix hierarchy without fine-tuning.  

%Scalars
We kept all the terms in the scalar potential and 
studied in detail 
the $125$ GeV Higgs Yukawa couplings to the SM quarks in detail and traced the source of their suppression to the cancellation between terms in $125$ GeV Higgs eigenvectors. This effect takes place regardless of the quark mass matrices and is not dependent on the chosen quark mass matrix texture. 
We also found that for all generation assignments the coupling between the SM quarks and exotic quarks with $125$ GeV Higgs are not suppressed by the $SU(3)_L$-breaking scale. The coupling of the third  generation and exotic quarks can be particularly large, even order one. This comes out naturally, as the Yukawa couplings affecting top mass have to be close to one, as the top mass is at the electroweak scale. Also the coupling between bottom quark and exotic down quarks can be significant, but not as large as in the case of top.
We also computed analytically all the other scalars and pseudo-scalar flavour violating couplings and found that they are insignificant, independently of parameter values in the scalar of quark sector. 

%Gauge boson mediators and SM and exotic mixing
We took fully into account the mixing between SM and exotic quarks and found that it affects the flavour violating couplings between the SM quarks with $Z$ and $X^0$, but not with $Z'$ and $125$ GeV Higgs.
We find that $Z'$ dominates neutral meson mixing mediation at $SU(3)_L$-breaking scale $\gtrsim 10$ TeV for all generation assignments. The $Z$ and $X^0$ bosons can give significant contribution to meson mixing at  $SU(3)_L$-breaking scale $\lesssim 10$ TeV  for $D$-meson and kaon if third generation is discriminated, for $B_s$-meson if first generation is discriminated and for $B_d$-meson in the case of discriminated second generation. 

%Bound on BSM scale

BSM scale can be as low as $\sim 10$ TeV from meson mixing and our numerical bench marks show that $15$ TeV passes the collider bounds. 
The general 331 model with $\beta=-1/\sqrt{3}$ 
 can therefore be probed at the LHC and especially at the currently discussed future colliders, but only if the third generation is the discriminated one. For discriminated  first and second generation the neutral meson mixing imposes $SU(3)_L$-breaking scale has to be too high for that. We conclude that if the heavier 331 gauge bosons are discovered in the future colliders, the third generation is the discriminated one.

\vspace{0.5cm}

{\bf Acknowledgements.}  
This work was supported by the Estonian Research Council grants  PRG803, PRG1677, and the 
CoE program TK202 "Fundamental Universe".
%KH acknowledges the H2020-MSCA-RICE-2014 grant no. 645722 (NonMinimalHiggs). SM acknowledges DST-SERB India, Core Researh Grant no. CRG/2022/003208.

\appendix

%%%%%%%%%%%%%%%%%%%%%%%%%%%%%%%%%%%%%%%%%%%%

%%%%%%%%%%%%%%%%%%%%%%%%%%%%%%%%%%%%%%%%%%%%%

\section{$Z\textrm{-}Z'$-mixing }
The neutral gauge boson mass matrix in basis $(W_3, W_8, B)$ is:
\be\label{neutral gauge boson mass matrix}
\left(\begin{array}{ccc}
\frac{g_3^2}{4}(v_\rho^2+{v_\eta}^2) & \frac{g_3^2}{4\sqrt{3}}(v_\rho^2-{v_\eta}^2) & -\frac{g_3 g_x}{6}(v_\rho^2+2{v_\eta}^2)\\
\frac{g_3^2}{4\sqrt{3}}(v_\rho^2-{v_\eta}^2) & \frac{g_3^2}{12}(4u^2+v_\rho^2+{v_\eta}^2) & \frac{g_3 g_x}{6\sqrt{3}}(2u^2-v_\rho^2+2{v_\eta}^2)\\
 -\frac{g_3 g_x}{6}(v_\rho^2+2{v_\eta}^2) & \frac{g_3 g_x}{6\sqrt{3}}(2u^2-v_\rho^2+2{v_\eta}^2) & \frac{g_x^2}{9}(v_\chi^2+v_\rho^2+4{v_\eta}^2)
\end{array}
\right).
\ee
It has one zero eigenvalue corresponding to photon.
The photon can be separated from the other states that will become $Z$ and $Z'$ by writing the above mass matrix in a basis that uses the following basis vectors:
\bea
A_\mu & = \sin\theta_W W_{3\mu}+\cos\theta_W \left(-\frac{\tan\theta_W}{\sqrt{3}}W_{8\mu}+\frac{g_3}{g_x}\tan\theta_W B_\mu\right),\\
\widetilde{Z}_\mu & = \cos\theta_W W_{3\mu}-\sin\theta_W \left(-\frac{\tan\theta_W}{\sqrt{3}}W_{8\mu}+\frac{g_3}{g_x}\tan\theta_W B_\mu\right),\\
\widetilde{Z}'_\mu & = -\frac{g_3}{g_x}\tan\theta_W W_{8\mu}-\frac{\tan\theta_W}{\sqrt{3}}B_\mu,
\eea
where $A_\mu$ is the photon mass eigenstate. The states $\widetilde{Z}_\mu$ and $\widetilde{Z}'_\mu$ are not mass eigenstates.

By writing the mass matrix \ref{neutral gauge boson mass matrix} in basis $(A, \widetilde{Z}, \widetilde{Z}')$, the photon drops out and we are left with a $2\times 2$-block in $(\widetilde{Z}, \widetilde{Z}')$-basis:
\be
M_{Z-Z'}^2=\tan^2\theta_W\left(\begin{array}{cc}
a & c\\
c & b
\end{array}
\right),
\ee
where
\begin{eqnarray}
a && =  \frac{g_3^2(v_\rho^2+{v_\eta}^2)}{4\sin^2\theta_W},\\
b && = \frac{\left[9g_3^4(4u^2+v_\rho^2+{v_\eta}^2)
+12g_3^2 g_x^2(2u^2-v_\rho^2+2{v_\eta}^2)
+4g_x^4(v_\chi^2+v_\rho^2+4{v_\eta}^2)\right]}{108g_x^2},\\
c && = -\frac{g_3\left[3g_3^2(v_\rho^2-{v_\eta}^2)-2g_x^2(v_\rho^2+2{v_\eta}^2)\right]}{12\sqrt{3}g_x\sin\theta_W}.
\end{eqnarray}
This is diagonalized as:
\be
U^T M_{Z-Z'}^2 U = \left(\begin{array}{cc}
m_Z^2 & 0\\
0 & m_{Z'}^2
\end{array}
\right),\quad \textrm{with} \quad
U=\left(\begin{array}{cc}
\cos\theta & \sin\theta\\
-\sin\theta & \cos\theta
\end{array}
\right),
\ee
where the $Z-Z'$ mixing angle is:
\be
\tan 2\theta = \frac{g_3\sqrt{3}\sin\theta_W\left[(2\sin^2\theta_W-1)v_\rho^2+{v_\eta}^2\right]}{g_x\left[2\cos^4\theta_W v_\chi^2+(2\sin^4\theta_W-1)v_\rho^2+(2\sin^2\theta_W-1){v_\eta}^2\right]}.
\ee

\section{Gauge boson currents}\label{gauge boson currents}

\subsection{Neutral currents}\label{gauge boson neutral currents}

The physical gauge boson couplings are 
\begin{eqnarray}
&&\mathcal{L}_{\rm gauge} 
=\bar u_L U_L^u{\lambda'}_Z^uU_L^{u\dagger}\gamma^\mu u_L Z_\mu
+\bar d_L U_L^d{\lambda'}_Z^dU_L^{d\dagger}\gamma^\mu d_L Z_\mu
+\bar u_L U_L^u{\lambda'}_{Z'}^uU_L^{u\dagger}\gamma^\mu u_L Z'_\mu\\
&&+
\bar d_L U_L^d{\lambda'}_{Z'}^d U_L^{d\dagger}\gamma^\mu d_L Z'_\mu
+(\bar u_L U_L^u{\lambda'}_{X^0}^u U_L^{u\dagger}\gamma^\mu u_L X^0_\mu
+\bar d_L U_L^d{\lambda'}_{X^0}^d U_L^{d\dagger}\gamma^\mu d_L X^0_\mu+\hc).\nonumber
\end{eqnarray}
The gauge boson couplings in Eq. \eqref{gauge boson physical couplings} are related to quark flavour rotation as: 
\be
{\lambda}_{Z,Z',X^0,\mathcal{G}}^q=U_L^q({\lambda'}_{Z,Z',X^0,\mathcal{G}}^q)U_L^{q\dagger}, \quad\quad q=u,d.
\ee
The non-rotated couplings matrices in the case of discriminated first generation for $Z$ and $Z'$ are
\be\label{Z and Zprime up before}
({\lambda'}_{Z,Z',1st}^u) = \left(\begin{array}{cccc}
a_{Z,Z'}^u +b_{Z,Z'}^u & 0 & 0 & 0\\
0 & a_{Z,Z'}^u & 0 & 0\\
0 & 0 & a_{Z,Z'}^u & 0\\
0 & 0 & 0 & a_{Z,Z'}^u+c_{Z,Z'}^u
\end{array}\right),
\ee
\be\label{Z and Zprime down before}
({\lambda'}_{Z,Z',1st}^d) = \left(\begin{array}{ccccc}
a_{Z,Z'}^d+b_{Z,Z'}^d & 0 & 0 & 0 & 0\\
0 & a_{Z,Z'}^d & 0 & 0 & 0\\
0 & 0 & a_{Z,Z'}^d & 0 & 0\\
0 & 0 & 0 & a_{Z,Z'}^d+c_{Z,Z'}^d & 0\\
0 & 0 & 0 & 0 & a_{Z,Z'}^d+c_{Z,Z'}^d
\end{array}\right),
\ee
where 
\begin{eqnarray}
a_Z^u & = & \frac{g_3^2\sin\theta_W(3 g_3\sin\theta_W\cos\theta-\sqrt{3}g_x\sin\theta)}{6g_x^2\cos\theta_W},\quad b_Z^u  = \frac{g_x\sin\theta}{\sqrt{3}\tan\theta_W},\\
c_Z^u & = & \frac{-9g_3 g_x\cos\theta+\sqrt{3}(-3g_3^2+2g_x^2)\sin\theta_W\sin\theta}{18g_x\cos\theta_W},\\
%%%%%%%%%%%%%%%%%%%%%
a_Z^d & = & -\frac{g_3 (g_x (1+2\cos^2\theta_W)\cos\theta+\sqrt{3}g_3 \sin\theta_W\sin\theta)}{6g_x\cos\theta_W},\quad b_Z^d  =  b_Z^u,\\
c_Z^d & = & \frac{g_3 (g_x \cos\theta+\sqrt{3}g_3 \sin\theta_W\sin\theta)}{2g_x\cos\theta_W},\\
%%%%%%%%%%%%%%%%%%%%%
a_{Z'}^u & = & \frac{g_3^2\sin\theta_W(3 g_3\sin\theta_W\sin\theta+\sqrt{3}g_x\cos\theta)}{6g_x^2\cos\theta_W},\quad b_{Z'}^u  =  -\frac{g_x\cos\theta}{\sqrt{3}\tan\theta_W},\\
c_{Z'}^u & = & \frac{-9g_3 g_x\sin\theta+\sqrt{3}(3g_3^2-2g_x^2)\sin\theta_W\cos\theta}{18g_x\cos\theta_W},\\
%%%%%%%%%%%%%%%%%%%%%
a_{Z'}^d & = & \frac{g_3 (-g_x (1+2\cos^2\theta_W)\sin\theta+\sqrt{3}g_3 \sin\theta_W\cos\theta)}{6g_x\cos\theta_W},\quad b_{Z'}^d  =  b_{Z'}^u,\\
c_{Z'}^d & = & \frac{g_3 (g_x \sin\theta-\sqrt{3}g_3 \sin\theta_W\cos\theta)}{2g_x\cos\theta_W}.
\end{eqnarray}
The non-rotated coupling matrices for discriminated second and third generation are obtained from Eqs. \eqref{Z and Zprime up before} and  \eqref{Z and Zprime down before} by moving term $b^q_{Z,Z'}$ to second and third diagonal term, respectively. 
In Eqs. \eqref{Z and Zprime up before} and  \eqref{Z and Zprime down before} we have separated the part proportional to unit matrix. As the quark rotation matrices will pass through $a_{Z,Z'}^u$ terms due to unitarity, only the terms $b_{Z,Z'}^u$ and $c_{Z,Z'}^u$ end up contributing to the off-diagonal couplings.

The $X^0$ non-rotated up-type coupling matrices $({\lambda'}_{X^0,\text{1st}}^u),~({\lambda'}_{X^0,\text{2nd}}^u),~({\lambda'}_{X^0,\text{3rd}}^u)$ are
\be\label{x u before rotation}
\frac{g_3}{\sqrt{2}}\left(\begin{array}{cccc}
0 & 0 & 0 & 1\\
0 & 0 & 0 & 0\\
0 & 0 & 0 & 0\\
0 & 0 & 0 & 0
\end{array}\right),\quad
\frac{g_3}{\sqrt{2}}\left(\begin{array}{cccc}
0 & 0 & 0 & 0\\
0 & 0 & 0 & 1\\
0 & 0 & 0 & 0\\
0 & 0 & 0 & 0
\end{array}\right),\quad
\frac{g_3}{\sqrt{2}}\left(\begin{array}{cccc}
0 & 0 & 0 & 0\\
0 & 0 & 0 & 0\\
0 & 0 & 0 & 1\\
0 & 0 & 0 & 0
\end{array}\right),
\ee
respectively and the $X^0$ non-rotated down-type coupling matrices $({\lambda'}_{X^0,\text{1st}}^d),~({\lambda'}_{X^0,\text{2nd}}^d)$, $({\lambda'}_{X^0,\text{3rd}}^d)$ are
\be\label{x down before rotation}
-\frac{g_3}{\sqrt{2}}\left(\begin{array}{ccccc}
0 & 0 & 0 & 0 & 0\\
0 & 0 & 0 & 0 & 0\\
0 & 0 & 0 & 0 & 0\\
0 & 1 & 0 & 0 & 0\\
0 & 0 & 1 & 0 & 0
\end{array}\right),\quad
-\frac{g_3}{\sqrt{2}}\left(\begin{array}{ccccc}
0 & 0 & 0 & 0 & 0\\
0 & 0 & 0 & 0 & 0\\
0 & 0 & 0 & 0 & 0\\
1 & 0 & 0 & 0 & 0\\
0 & 0 & 1 & 0 & 0
\end{array}\right),\quad
-\frac{g_3}{\sqrt{2}}\left(\begin{array}{ccccc}
0 & 0 & 0 & 0 & 0\\
0 & 0 & 0 & 0 & 0\\
0 & 0 & 0 & 0 & 0\\
1 & 0 & 0 & 0 & 0\\
0 & 1 & 0 & 0 & 0
\end{array}\right),
\ee
respectively.
As the non-zero elements are located off-diagonal, the resulting physical couplings after flavour rotation will have large flavour violating couplings.

\section{Numerical examples}\label{numerical example}

The mass matrices used in the benchmarks are parametrized as:
\be
m_u=\left(
\begin{array}{cccc}
-v_\eta c^u_{11}\epsilon^{8} &
-v_\eta c^u_{12}\epsilon^{5} & 
-v_\eta c^u_{13}\epsilon^{3} & 
-v_\eta c^u_{14}\epsilon^{3}\\
-v_\eta c^u_{21}\epsilon^{7} &
-v_\eta c^u_{22}\epsilon^{4} & 
-v_\eta c^u_{23}\epsilon^{2} &
-v_\eta c^u_{24}\epsilon^{2} \\
 v_\rho c^u_{31}\epsilon^{5} &
 v_\rho c^u_{32}\epsilon^{2} & 
 v_\rho c^u_{33}\epsilon^{0} &
 v_\rho c^u_{34}\epsilon^{0}\\
v_\chi c^u_{41}\epsilon^{7} & 
v_\chi c^u_{42}\epsilon^{2} & 
v_\chi c^u_{43}\epsilon^{2} & 
v_\chi c^u_{44}\epsilon^{2}\\
\end{array}
\right),
\ee
\be
m_d=
\left(
\begin{array}{ccccc}
v_\rho  c^d_{11}\epsilon^{7} &
v_\rho  c^d_{11}\epsilon^{6} &
v_\rho  c^d_{11}\epsilon^{6} &
v_\rho  c^d_{11}\epsilon^{5} &
v_\rho  c^d_{11}\epsilon^{4}  \\
v_\rho  c^d_{11}\epsilon^{6} &
v_\rho  c^d_{11}\epsilon^{5} &
v_\rho  c^d_{11}\epsilon^{5} &
v_\rho  c^d_{11}\epsilon^{4} &
v_\rho  c^d_{11}\epsilon^{3}  \\
v_\eta c^d_{11}\epsilon^{4} &
v_\eta c^d_{11}\epsilon^{3} & 
v_\eta c^d_{11}\epsilon^{3} &
v_\eta c^d_{11}\epsilon^{2} &
v_\eta c^d_{11}\epsilon^{1} \\
v_\chi c^d_{11}\epsilon^{3} &
v_\chi c^d_{11}\epsilon^{2} & 
v_\chi c^d_{11}\epsilon^{2} &
v_\chi c^d_{11}\epsilon^{1} &
v_\chi c^d_{11}\epsilon^{0} \\
v_\chi c^d_{11}\epsilon^{3} &
v_\chi c^d_{11}\epsilon^{2} & 
v_\chi c^d_{11}\epsilon^{2} &
v_\chi c^d_{11}\epsilon^{1} &
v_\chi c^d_{11}\epsilon^{0}  
\end{array}
\right).
\ee
For all three benchmarks we set $v_\rho = v_\eta = 246/\sqrt{2}\GeV$  and $v_\chi=$15 TeV.

\subsection{Benchmark point 1}\label{BP1}
The used order-one couplings of Eqs. (\ref{up mass matrix}) and (\ref{down mass matrix})  are:

\be
c^u=\left(\begin{array}{cccc}
4.97073 & 2.7245 & 4.01503 & 1.67706\\
5.87\times 10^{-11} & 4.46721 & 3.50154 & 2.32508\\
20.2741 & 3.78\times 10^{-14} & 2.71875 & 2.17965\\
1.76\times 10^{-10} & 2.08492 & 2.25081 & 0.56448\\
\end{array}\right),
\ee
\be
c^d=\left(\begin{array}{ccccc}
1.75467 & 1.68135 & 1.66029 & 2.60774 & 2.1004\\
1.18195 & 3.50899 & 4.06419 & 4.20559 & 3.5717\\
0.000107515 & 1.988 & 5.62861 & 4.19889 & 3.08267\\
2.55278 & 4.9599 & 3.28512 & 0.0314777 & 0.682893\\
1.61006 & 3.29462 & 2.93614 & 1.09371 & 1.27827\\
\end{array}\right).
\ee
The left-handed rotation matrices are
\be
U_L^u=\left(\begin{array}{cccc}
1.0 & -0.0085 & 0.0050 & 0.0033 \\
-0.0082 & 1.0 & -0.050 & -0.0048 \\
0.0042 & 0.047 & -0.96 & 0.29\\
-0.0048 & -0.019 & 0.29 & 0.96 \\
\end{array}\right),
\ee
\be
U_L^d=\left(\begin{array}{ccccc}
0.97 & -0.24 & 0.00023 & -0.000021 & 0.000051\\
-0.24 & -0.97 & -0.0053 & -0.00031 & 0.00060\\
0.0015 & 0.0052 & -1.0 & -0.0054 & 0.0094\\
0.000076 & 0.00051 & 0.0092 & -0.87 & 0.48\\
0.000046 & 0.00034 & 0.0056 & 0.48 & 0.87\\
\end{array}\right).
\ee

\subsection{Benchmark point 2}\label{BP2}

\be
c^u=\left(\begin{array}{cccc}
31.7597 & 3.08122 & 4.39011 & 1.44231\\
9.43\times 10^{-7} & 2.83497 & 2.19622 & 2.85768\\
135.587 & 6.3701 & 1.81001 & 2.72798\\
0.0000741971 & 2.54977 & 3.56919 & 2.1502\\
\end{array}\right),
\ee
\be
c^d=\left(\begin{array}{ccccc}
2.75324 & 1.34124 & 2.52898 & 2.26037 & 1.46415\\
3.14726 & 0.694873 & 3.7127 & 4.30212 & 1.8983\\
1.31444 & 0.0522147 & 3.00486 & 0.128547 & 0.223748\\
5.2971 & 2.41595 & 3.84795 & 4.1146 & 3.26315\\
6.10452 & 0.874773 & 2.86149 & 3.12481 & 0.178137\\
\end{array}\right).
\ee
The left-handed rotation matrices are
\be
U_L^u=\left(\begin{array}{cccc}
-0.97 & -0.23 & -0.0029 & -0.0045\\
0.23 & -0.97 & -0.049 & -0.00077\\
-0.0092 & 0.047 & -0.99 & 0.16 \\
-0.0028 & -0.0093 & 0.16 & 0.99\\
\end{array}\right),
\ee
\be
U_L^d=\left(\begin{array}{ccccc}
1.0 & 0.0079 & -0.0017 & -0.000015 & -5.6\times 10^{-6}\\
-0.0079 & 1.0 & 0.0054 & -0.000078 & -0.000091\\
-0.0017 & 0.0054 & -1.0 & 0.00018 & -0.000015\\
3.2\times 10^{-6} & 0.000081 & -0.000036 & -0.12 & 0.99\\
-0.000015 & -0.000087 & 0.00018 & -0.99 & -0.12\\
\end{array}\right).
\ee

\subsection{Benchmark point 3}\label{BP3}

\be
c^u=\left(\begin{array}{cccc}
0.1 & 0.75271 & 4.80535 & 0.552645\\
4.00015 & 4.45161 & 0.713134 & 1.34193\\
7.4217 & 2.89162 & 2.54457 & 2.40413\\
0.389465 & 3.35982 & 3.00134 & 1.04204\\
\end{array}\right),
\ee
\be
c^d=\left(\begin{array}{ccccc}
2.12673 & 3.43606 & 0.324718 & 3.19138 & 1.1927\\
4.19815 & 1.28651 & 0.538586 & 3.50963 & 4.9621\\
0.115547 & 2.35043 & 2.84841 & 2.55706 & 0.0722823\\
2.92591 & 0.378945 & 0.189045 & 0.773807 & 2.23766\\
2.62046 & 4.31457 & 0.266703 & 4.46023 & 1.0141\\
\end{array}\right).
\ee
The left-handed rotation matrices are
\be
U_L^u=\left(\begin{array}{cccc}
0.96 & 0.28 & 0.0021 & 0.0045\\
0.28 & -0.96 & -0.039 & 0.0057\\
-0.0010 & 0.038 & -0.96 & 0.22\\
-0.0038 & -0.0041 & 0.22 & 0.98\\
\end{array}\right),
\ee
\be
U_L^d=\left(\begin{array}{ccccc}
1.0 & 0.056 & -0.00052 & -0.000024 & -0.000025\\
0.056 & -1.0 & 0.0036 & 0.00029 & 0.000052\\
-0.00032 & 0.0036 & -1.0 & -0.00066 & 0.0016\\
0.000016 & -0.000090 & 0.0018 & -0.50 & 0.87\\
-0.000017 & -0.00028 & -0.00024 & -0.87 & -0.50\\
\end{array}\right).
\ee

\bibliographystyle{JHEP}
\bibliography{collider_331}

\providecommand{\href}[2]{#2}\begingroup\raggedright\begin{thebibliography}{10}

\bibitem{Deppisch:2016jzl}
F.~F. Deppisch, C.~Hati, S.~Patra, U.~Sarkar, and J.~W.~F. Valle, {\it {331
  Models and Grand Unification: From Minimal SU(5) to Minimal SU(6)}},  {\em
  Phys. Lett. B} {\bf 762} (2016) 432--440,
  [\href{http://arxiv.org/abs/1608.05334}{{\tt arXiv:1608.05334}}].

\bibitem{Sanchez:2001ua}
L.~A. Sanchez, W.~A. Ponce, and R.~Martinez, {\it {SU(3) ($c$) x SU(3) ($\ell$)
  x U(1) ($X$) as an E(6) subgroup}},  {\em Phys. Rev. D} {\bf 64} (2001)
  075013, [\href{http://arxiv.org/abs/hep-ph/0103244}{{\tt hep-ph/0103244}}].

\bibitem{Glashow:1976nt}
S.~L. Glashow and S.~Weinberg, {\it {Natural Conservation Laws for Neutral
  Currents}},  {\em Phys. Rev. D} {\bf 15} (1977) 1958.

\bibitem{Paschos:1976ay}
E.~A. Paschos, {\it {Diagonal Neutral Currents}},  {\em Phys. Rev. D} {\bf 15}
  (1977) 1966.

\bibitem{Frampton:1994rt}
P.~H. Frampton, {\it {The Third family is different}},  in {\em {Particles,
  Strings, and Cosmology (PASCOS 94)}}, pp.~0063--80, 9, 1994.
\newblock \href{http://arxiv.org/abs/hep-ph/9409331}{{\tt hep-ph/9409331}}.

\bibitem{GomezDumm:1993oxo}
D.~Gomez~Dumm, F.~Pisano, and V.~Pleitez, {\it {Flavor changing neutral
  currents in SU(3) x U(1) models}},  {\em Mod. Phys. Lett. A} {\bf 9} (1994)
  1609--1615, [\href{http://arxiv.org/abs/hep-ph/9307265}{{\tt
  hep-ph/9307265}}].

\bibitem{Liu:1994rx}
J.~T. Liu, {\it {Generation nonuniversality and flavor changing neutral
  currents in the 331 model}},  {\em Phys. Rev. D} {\bf 50} (1994) 542--547,
  [\href{http://arxiv.org/abs/hep-ph/9312312}{{\tt hep-ph/9312312}}].

\bibitem{Long:1999ij}
H.~N. Long and V.~T. Van, {\it {Quark family discrimination and flavor changing
  neutral currents in the SU(3)(C) x SU(3)(L) x U(1) model with right-handed
  neutrinos}},  {\em J. Phys. G} {\bf 25} (1999) 2319--2324,
  [\href{http://arxiv.org/abs/hep-ph/9909302}{{\tt hep-ph/9909302}}].

\bibitem{Rodriguez:2004mw}
J.~A. Rodriguez and M.~Sher, {\it {FCNC and rare B decays in 3-3-1 models}},
  {\em Phys. Rev. D} {\bf 70} (2004) 117702,
  [\href{http://arxiv.org/abs/hep-ph/0407248}{{\tt hep-ph/0407248}}].

\bibitem{Oliveira:2022vjo}
V.~Oliveira and C.~A.~d. S.~Pires, {\it {Flavor Changing Neutral Current
  processes and family discrimination in 3-3-1 models}},
  \href{http://arxiv.org/abs/2208.00420}{{\tt arXiv:2208.00420}}.

\bibitem{Promberger:2007py}
C.~Promberger, S.~Schatt, and F.~Schwab, {\it {Flavor Changing Neutral Current
  Effects and CP Violation in the Minimal 3-3-1 Model}},  {\em Phys. Rev. D}
  {\bf 75} (2007) 115007, [\href{http://arxiv.org/abs/hep-ph/0702169}{{\tt
  hep-ph/0702169}}].

\bibitem{Fritzsch:1977vd}
H.~Fritzsch, {\it {Weak Interaction Mixing in the Six - Quark Theory}},  {\em
  Phys. Lett. B} {\bf 73} (1978) 317--322.

\bibitem{Fritzsch:1979zq}
H.~Fritzsch, {\it {Quark Masses and Flavor Mixing}},  {\em Nucl. Phys. B} {\bf
  155} (1979) 189--207.

\bibitem{Cabarcas:2007my}
J.~M. Cabarcas, D.~Gomez~Dumm, and R.~Martinez, {\it {Constraints on economical
  331 models from mixing of $K$, Bd and $B_s$ neutral mesons}},  {\em Phys.
  Rev. D} {\bf 77} (2008) 036002, [\href{http://arxiv.org/abs/0711.2467}{{\tt
  arXiv:0711.2467}}].

\bibitem{Martinez:2008jj}
R.~Martinez and F.~Ochoa, {\it {Mass-matrix ansatz and constraints on B0(s) -
  anti-B0(s) mixing in 331 models}},  {\em Phys. Rev. D} {\bf 77} (2008)
  065012, [\href{http://arxiv.org/abs/0802.0309}{{\tt arXiv:0802.0309}}].

\bibitem{Belfatto:2023qca}
B.~Belfatto and Z.~Berezhiani, {\it {Minimally modified Fritzsch texture for
  quark masses and CKM mixing}},  \href{http://arxiv.org/abs/2305.00069}{{\tt
  arXiv:2305.00069}}.

\bibitem{CarcamoHernandez:2005ka}
A.~E. Carcamo~Hernandez, R.~Martinez, and F.~Ochoa, {\it {Z and Z' decays with
  and without FCNC in 331 models}},  {\em Phys. Rev. D} {\bf 73} (2006) 035007,
  [\href{http://arxiv.org/abs/hep-ph/0510421}{{\tt hep-ph/0510421}}].

\bibitem{Georgi:1978bv}
H.~Georgi and A.~Pais, {\it {Generalization of Gim: Horizontal and Vertical
  Flavor Mixing}},  {\em Phys. Rev. D} {\bf 19} (1979) 2746.

\bibitem{Singer:1980sw}
M.~Singer, J.~W.~F. Valle, and J.~Schechter, {\it {Canonical Neutral Current
  Predictions From the Weak Electromagnetic Gauge Group SU(3) X $u$(1)}},  {\em
  Phys. Rev. D} {\bf 22} (1980) 738.

\bibitem{Valle:1983dk}
J.~W.~F. Valle and M.~Singer, {\it {Lepton Number Violation With Quasi Dirac
  Neutrinos}},  {\em Phys. Rev. D} {\bf 28} (1983) 540.

\bibitem{Montero:1992jk}
J.~C. Montero, F.~Pisano, and V.~Pleitez, {\it {Neutral currents and GIM
  mechanism in SU(3)-L x U(1)-N models for electroweak interactions}},  {\em
  Phys. Rev. D} {\bf 47} (1993) 2918--2929,
  [\href{http://arxiv.org/abs/hep-ph/9212271}{{\tt hep-ph/9212271}}].

\bibitem{Foot:1994ym}
R.~Foot, H.~N. Long, and T.~A. Tran, {\it {$SU(3)_L \otimes U(1)_N$ and
  $SU(4)_L \otimes U(1)_N$ gauge models with right-handed neutrinos}},  {\em
  Phys. Rev. D} {\bf 50} (1994), no.~1 R34--R38,
  [\href{http://arxiv.org/abs/hep-ph/9402243}{{\tt hep-ph/9402243}}].

\bibitem{Long:1995ctv}
H.~N. Long, {\it {The 331 model with right handed neutrinos}},  {\em Phys. Rev.
  D} {\bf 53} (1996) 437--445, [\href{http://arxiv.org/abs/hep-ph/9504274}{{\tt
  hep-ph/9504274}}].

\bibitem{Long:1996rfd}
H.~N. Long, {\it {SU(3)-L x U(1)-N model for right-handed neutrino neutral
  currents}},  {\em Phys. Rev. D} {\bf 54} (1996) 4691--4693,
  [\href{http://arxiv.org/abs/hep-ph/9607439}{{\tt hep-ph/9607439}}].

\bibitem{Pleitez:1994pu}
V.~Pleitez, {\it {New fermions and a vector - like third generation in SU(3)
  (C) x SU(3) (L) x U(1) ($N$) models}},  {\em Phys. Rev. D} {\bf 53} (1996)
  514--526, [\href{http://arxiv.org/abs/hep-ph/9412304}{{\tt hep-ph/9412304}}].

\bibitem{Long:1997vbr}
H.~N. Long, {\it {Scalar sector of the 3 3 1 model with three Higgs triplets}},
   {\em Mod. Phys. Lett. A} {\bf 13} (1998) 1865--1874,
  [\href{http://arxiv.org/abs/hep-ph/9711204}{{\tt hep-ph/9711204}}].

\bibitem{Dong:2013ioa}
P.~V. Dong, T.~P. Nguyen, and D.~V. Soa, {\it {3-3-1 model with inert scalar
  triplet}},  {\em Phys. Rev. D} {\bf 88} (2013), no.~9 095014,
  [\href{http://arxiv.org/abs/1308.4097}{{\tt arXiv:1308.4097}}].

\bibitem{Pisano:1992bxx}
F.~Pisano and V.~Pleitez, {\it {An SU(3) x U(1) model for electroweak
  interactions}},  {\em Phys. Rev. D} {\bf 46} (1992) 410--417,
  [\href{http://arxiv.org/abs/hep-ph/9206242}{{\tt hep-ph/9206242}}].

\bibitem{Frampton:1992wt}
P.~H. Frampton, {\it {Chiral dilepton model and the flavor question}},  {\em
  Phys. Rev. Lett.} {\bf 69} (1992) 2889--2891.

\bibitem{Foot:1992rh}
R.~Foot, O.~F. Hernandez, F.~Pisano, and V.~Pleitez, {\it {Lepton masses in an
  SU(3)-L x U(1)-N gauge model}},  {\em Phys. Rev. D} {\bf 47} (1993)
  4158--4161, [\href{http://arxiv.org/abs/hep-ph/9207264}{{\tt
  hep-ph/9207264}}].

\bibitem{Tonasse:1996cx}
M.~D. Tonasse, {\it {The Scalar sector of 3-3-1 models}},  {\em Phys. Lett. B}
  {\bf 381} (1996) 191--201, [\href{http://arxiv.org/abs/hep-ph/9605230}{{\tt
  hep-ph/9605230}}].

\bibitem{Nguyen:1998ui}
T.~A. Nguyen, N.~A. Ky, and H.~N. Long, {\it {The Higgs sector of the minimal 3
  3 1 model revisited}},  {\em Int. J. Mod. Phys. A} {\bf 15} (2000) 283--305,
  [\href{http://arxiv.org/abs/hep-ph/9810273}{{\tt hep-ph/9810273}}].

\bibitem{Benavides:2009cn}
R.~H. Benavides, Y.~Giraldo, and W.~A. Ponce, {\it {FCNC in the 3-3-1 model
  with right-handed neutrinos}},  {\em Phys. Rev. D} {\bf 80} (2009) 113009,
  [\href{http://arxiv.org/abs/0911.3568}{{\tt arXiv:0911.3568}}].

\bibitem{Froggatt:1978nt}
C.~D. Froggatt and H.~B. Nielsen, {\it {Hierarchy of Quark Masses, Cabibbo
  Angles and CP Violation}},  {\em Nucl. Phys. B} {\bf 147} (1979) 277--298.

\bibitem{Huitu:2019mdr}
K.~Huitu, N.~Koivunen, and T.~J. K\"arkk\"ainen, {\it {Natural neutrino sector
  in a 331-model with Froggatt-Nielsen mechanism}},  {\em JHEP} {\bf 02} (2020)
  162, [\href{http://arxiv.org/abs/1908.09384}{{\tt arXiv:1908.09384}}].

\bibitem{ParticleDataGroup:2022pth}
{\bf Particle Data Group} Collaboration, R.~L. Workman et~al., {\it {Review of
  Particle Physics}},  {\em PTEP} {\bf 2022} (2022) 083C01.

\bibitem{Wang:2019try}
B.~Wang, {\it {Calculation of the $K_L-K_S$ mass difference for physical quark
  masses}},  {\em PoS} {\bf LATTICE2019} (2019) 093,
  [\href{http://arxiv.org/abs/2001.06374}{{\tt arXiv:2001.06374}}].

\bibitem{Wang:2022lfq}
B.~Wang, {\it {Calculating $\Delta m_K$ with lattice QCD}},  {\em PoS} {\bf
  LATTICE2021} (2022) 141, [\href{http://arxiv.org/abs/2301.01387}{{\tt
  arXiv:2301.01387}}].

\bibitem{DeBruyn:2022zhw}
K.~De~Bruyn, R.~Fleischer, E.~Malami, and P.~van Vliet, {\it {New physics in
  B$_{q}$ $^{0}$\textendash{} mixing: present challenges, prospects, and
  implications for}},  {\em J. Phys. G} {\bf 50} (2023), no.~4 045003,
  [\href{http://arxiv.org/abs/2208.14910}{{\tt arXiv:2208.14910}}].

\bibitem{Gabbiani:1996hi}
F.~Gabbiani, E.~Gabrielli, A.~Masiero, and L.~Silvestrini, {\it {A Complete
  analysis of FCNC and CP constraints in general SUSY extensions of the
  standard model}},  {\em Nucl. Phys. B} {\bf 477} (1996) 321--352,
  [\href{http://arxiv.org/abs/hep-ph/9604387}{{\tt hep-ph/9604387}}].

\bibitem{Huitu:2019kbm}
K.~Huitu and N.~Koivunen, {\it {Suppression of scalar mediated FCNCs in a
  $SU(3)_c\times SU(3)_L\times U(1)_X$-model}},  {\em JHEP} {\bf 10} (2019)
  065, [\href{http://arxiv.org/abs/1905.05278}{{\tt arXiv:1905.05278}}].

\bibitem{CMS:2022fck}
{\bf CMS} Collaboration, A.~Tumasyan et~al., {\it {Search for pair production
  of vector-like quarks in leptonic final states in proton-proton collisions at
  $ \sqrt{s} $ = 13 TeV}},  {\em JHEP} {\bf 07} (2023) 020,
  [\href{http://arxiv.org/abs/2209.07327}{{\tt arXiv:2209.07327}}].

\bibitem{ATLAS:2024gyc}
{\bf ATLAS} Collaboration, G.~Aad et~al., {\it {Search for pair-production of
  vector-like quarks in lepton+jets final states containing at least one
  b-tagged jet using the Run 2 data from the ATLAS experiment}},  {\em Phys.
  Lett. B} {\bf 854} (2024) 138743,
  [\href{http://arxiv.org/abs/2401.17165}{{\tt arXiv:2401.17165}}].

\bibitem{CMS:2022tdo}
{\bf CMS} Collaboration, A.~Tumasyan et~al., {\it {Search for a W' boson
  decaying to a vector-like quark and a top or bottom quark in the all-jets
  final state at $ \sqrt{\mathrm{s}} $ = 13 TeV}},  {\em JHEP} {\bf 09} (2022)
  088, [\href{http://arxiv.org/abs/2202.12988}{{\tt arXiv:2202.12988}}].

\bibitem{CMS:2017hwg}
{\bf CMS} Collaboration, A.~M. Sirunyan et~al., {\it {Search for a heavy
  resonance decaying to a top quark and a vector-like top quark at $
  \sqrt{s}=13 $ TeV}},  {\em JHEP} {\bf 09} (2017) 053,
  [\href{http://arxiv.org/abs/1703.06352}{{\tt arXiv:1703.06352}}].

\bibitem{CMS:2018wkq}
{\bf CMS} Collaboration, A.~M. Sirunyan et~al., {\it {Search for a heavy
  resonance decaying to a top quark and a vector-like top quark in the
  lepton+jets final state in pp collisions at $\sqrt{s} =$ 13 TeV}},  {\em Eur.
  Phys. J. C} {\bf 79} (2019), no.~3 208,
  [\href{http://arxiv.org/abs/1812.06489}{{\tt arXiv:1812.06489}}].

\bibitem{CMS:2024lwn}
{\bf CMS} Collaboration, {\it {Model-agnostic search for dijet resonances with
  anomalous jet substructure in proton-proton collisions at $\sqrt{s}$ = 13
  TeV}}, .

\bibitem{ATLAS:2018tvr}
{\bf ATLAS} Collaboration, {\it {Prospects for searches for heavy $Z^\prime$
  and $W^\prime$ bosons in fermionic final states with the ATLAS experiment at
  the HL-LHC}}, .

\end{thebibliography}\endgroup

\end{document}